\documentclass[12pt,onecolumn,draftclsnofoot]{IEEEtran}
\usepackage{amssymb,amsmath,setspace,mathbbol}
\usepackage{algorithmicx,algorithm,graphicx,subfigure,epstopdf,url}
\usepackage{cite,color,hyperref}
\newtheorem{remark}{\textbf{Remark}}
\newtheorem{theorem}{\textbf{Theorem}}
\newtheorem{lemma}{\textbf{Lemma}}
\newtheorem{corollary}{\textbf{Corollary}}

\begin{document}
\title{On Jamming Against Wireless Networks\thanks{The authors are affiliated with Wireless @ VT, Bradley Department of Electrical and Computer Engineering, Virginia Tech. Email: \{adhiraj, hdhillon, rbuehrer\}@vt.edu. \hspace{140pt}Manuscript last updated: September 30, 2015.}}
\author{SaiDhiraj Amuru, Harpreet S. Dhillon, R. Michael Buehrer\vspace{-40pt}}
\date{}
\maketitle
\noindent 
\begin{abstract}
In this paper, we study jamming attacks against wireless networks. Specifically, we consider a network of base stations (BS) or access points (AP) and investigate the impact of a fixed number of jammers that are randomly deployed according to a Binomial point process. We shed light on the network performance in terms of a) the outage probability and b) the error probability of a victim receiver in the downlink of this wireless network. We derive analytical expressions for both these metrics and discuss in detail how the jammer network must adapt to the various wireless network parameters in order to effectively attack the victim receivers. For instance, we will show that with only $1$ jammer per BS/AP a) the outage probability of the wireless network can be increased from $1\%$ (as seen in the non-jamming case) to $80\%$ and b) when retransmissions are used, the jammers cause the effective network activity factor (and hence the interference among the BSs) to be doubled. Furthermore, we show that the behavior of the jammer network as a function of the BS/AP density is not obvious. In particular, an interesting concave-type behavior is seen which indicates that the number of jammers required to attack the wireless network must scale with the BS density only until a certain value beyond which it decreases. In the context of error probability of the victim receiver, we study whether or not some recent results related to jamming in the point-to-point link scenario can be extended to the case of jamming against wireless networks. Numerical results are presented to validate the theoretical inferences presented. 
\end{abstract}
\begin{IEEEkeywords}
Networks, jamming, outage, error probability, Binomial point process, stochastic geometry. 
\end{IEEEkeywords}
\section{Introduction}
The inherent openness of the wireless medium makes it susceptible to both intentional and un-intentional interference. Interference from the neighboring cells is one of the major causes of un-intentional interference. On the flip side, intentional interference, such as jamming attacks, corresponds to adversarial attacks on a victim receiver. In this paper, we study jamming attacks against wireless networks. Jamming-related studies not only provide insights into offensive attack strategies, but also enable us to understand the vulnerabilities of existing systems. Most studies that are related to jamming attacks in the physical layer only consider the presence of a single node (source-destination pair) and develop optimal jamming strategies, see \cite{GlobecomJamming}-\hspace{-0.5pt}\cite{Taxonomy} and references therein. Having gained insights about the jamming behavior in a single communication link scenario, the next step is to understand the jamming behavior against networks. 

Jamming against wireless multi-hop networks has previously been addressed from an optimization perspective in \cite{JamPlac1}-\hspace{-0.5pt}\cite{FlowJamming}. More specifically, \cite{JamPlac1}-\hspace{-0.5pt}\cite{JamPlac3} consider the problem of jammer placement against wireless multi-hop
networks with the aim of stopping the flows in the network. The jammer-to-flow assignment problem i.e., optimally assigning jammers to stop flows in the network based on their locations and other constraints such as power, was considered in \cite{FlowJamming}. All these works model networks as a graph and study the jamming problem with an aim to find the best set of nodes/edges to attack so that the network is disconnected. While these studies indicate which nodes/links to be attacked, they do not address the problem of jamming attacks against wireless networks from a physical layer perspective and don't consider infrastructure networks such as cellular or WiFi. Therefore, in this paper, we address the problem of attacking a wireless network when the jammers are randomly deployed in a given area and how this attack can be realized at the physical layer. Since the victim receiver location is typically unknown \emph{a priori}, the jammers are randomly deployed in a given area. 

Specifically, in this paper we consider a wireless network comprising of BS or APs that are deployed in an area of interest and investigate the impact of a fixed number of jamming nodes that are randomly deployed in the same area. The network performance under the jamming attack is analyzed from the perspective of the downlink of a victim receiver that is accessing this wireless network. Since the number of jammer nodes is fixed, we model the jammer network using a Binomial point process (BPP) \cite{Haenggi_BPP}. Notice that such an analysis is practical and useful as it will help in understanding the number of jamming nodes that would be needed to disrupt a wireless network, for example to attack an enemy network in a military setting.

Using the proposed jammer network model, we analyze the jamming performance against the wireless network in terms of a) the outage probability and b) the error probability of a victim receiver in this wireless network. We derive analytical expressions for both these metrics and analyze in detail the jamming impact against the wireless network in the presence of shadowing and fading typically seen in wireless environments. As will be shown, accounting for log-normal shadowing in both the outage and error probability analysis is complicated. Therefore, in this paper, we use the Gaussian-Hermite quadrature approximation \cite{Abramovitz} to account for this log-normal shadowing and give simple analytical expressions for both these metrics. We also show how this approximation affects the theoretical outage and error probabilities by comparing them with numerical simulations. 

Using the closed form outage-probability expressions, we discuss in detail how the jammer network parameters must change according to the wireless network so as to maximize outages at the victim receiver. While the outage probability characterizes the overall network performance in terms of the interference experienced at the victim receiver, it typically ignores the signaling schemes used by the BS/APs to communicate with the victim receiver. For instance, irrespective of the modulation schemes employed by the various nodes in the network, the outage probability analysis uses the power levels of the various signals in order to evaluate the signal-to-interference and noise ratio at the victim receiver. Hence, the outage probability analysis alone does not cover all the aspects of jamming a wireless network. Therefore, in this paper we also study the jamming performance in terms of the symbol error probability of the victim receiver which explicitly takes into consideration the signaling schemes used by the BS/AP and also the jammers. 

The error probability of the victim receiver can be analyzed by using tools from stochastic geometry when random spatial distributions are considered \cite{MoeWin}. However, there is relatively limited work in the literature (related to non-jamming scenarios) that performs such an analysis (see \cite{DiRenzo}-\hspace{-0.5pt}\cite{DiRenzo_TCOM} and references therein). In \cite{DiRenzo}, \cite{ElSawy}, a novel approach termed as \emph{equivalent-in-distribution} is developed to estimate the error probability of the victim receiver in the presence of interference generated from a Poisson field which is modeled as an alpha-stable distribution. However, no such equivalent distributions or approximations are known for the cases in which the interference arises from a Binomial field \cite{Haenggi_BPP}, which is the model considered for the jammers in this paper. In \cite{DiRenzo_TCOM}, a nearest neighbor approximation (corresponding to the modulation scheme of the victim) is used to analyze the error probability in a non-jamming scenario. In this method, the error probability depends on the minimum distance of the modulation scheme and the number of nearest neighbors for the points in the modulation scheme. Since no equivalent distributions or approximations are known for the interference originating from a Binomial field, in this paper we follow the approach in \cite{DiRenzo_TCOM} to analyze the error probability of the victim receiver.  

The various parameters that impact the victim receiver performance include the shadowing variance, transmit power of the BSs, the BS density, the network loading, i.e., the density of simultaneously active BSs, the jammer signal power, and the number of jammers. The effect of each of these parameters on the jamming impact at the victim receiver is examined both analytically and numerically. For instance, we show that in order to maintain a constant outage probability at the victim receiver, the number of jammers a) increases with the victim signal power levels, b) decreases with increasing network load, and c) decreases with increased shadowing levels. The jamming impact against the wireless network is also addressed when retransmissions are employed. Furthermore, we will show that the jammer network obeys an interesting concave-type behavior as a function of the BS/AP density in order to maintain a constant outage probability at the victim receiver. Similar analysis of the jammer network impact is performed in the context of the error probability of the victim receiver. In addition to the above analysis, we discuss the impact of some recent findings related to jamming in a point-to-point link scenario \footnote{In \cite{ModulationJamming_Journal}, we showed that the optimal jamming signaling scheme against a digital amplitude-phase modulation scheme is not Gaussian signaling and that it depends on the victim signal parameters}, when analyzing jamming against wireless networks. While extending the analysis in \cite{ModulationJamming_Journal} to the case of networks is beyond the scope of this paper, we discuss in detail the behavior of various jamming signals using extensive simulations.

\section{System Model}\label{sec:SystemModel}
\begin{figure}[ht]
\vspace{-20pt}
\centering
\includegraphics[width=0.65\textwidth]{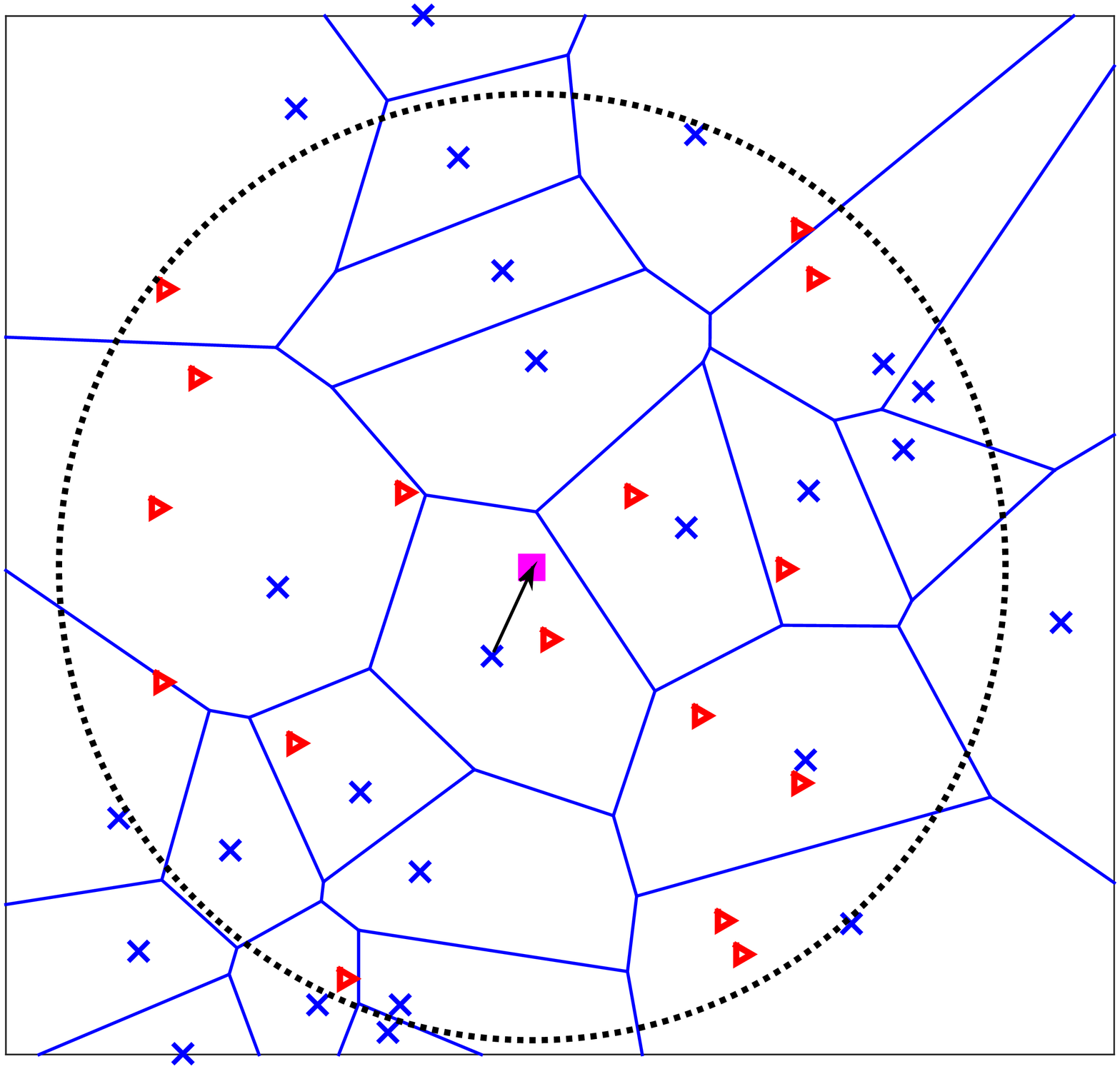}
\vspace{-20pt}
\caption{[System Model] The cross marks indicate the BS/APs in the wireless network that are distributed according to a PPP. The Voronoi tessellation indicates the coverage regions of the BS/APs. The square indicates the victim receiver which is at the origin.  The black arrow indicates the link between the the closest BS and the victim receiver. The triangles indicate the jammers that are distributed according to a BPP within the black-dotted region of radius $R_J$.}
\label{figSystemModel}
\end{figure}

We consider a victim wireless network of BSs or APs that are modeled according to a homogeneous Poisson point process (PPP) $\boldsymbol{\boldsymbol{\Psi}}$ of density $\lambda_T$ on $\mathbb{R}^2$ \cite{Andrews_HetNet}, \cite{PPP_Dhillon}. The downlink analysis in this paper is performed at the victim receiver which is assumed to be at the origin. The behavior of this wireless network is studied when it is attacked by a jammer network with $N_J$ jammers. We model the jammer network according to a BPP denoted by $\boldsymbol{\Psi}_J$. Fig.~\ref{figSystemModel} shows the wireless network, the victim receiver and the jammers that attack this wireless network. 

The received signal at the victim receiver is given by 
\begin{align}\label{eq:signal_model}
y=\sqrt{P_T\chi_0}(1+r_0)^{-\frac{\alpha}{2}}h_0s_0+\underbrace{\sum_{i\in\boldsymbol{\boldsymbol{\Psi}}\backslash \{0\}}a_i\sqrt{P_T\chi_i}(1+r_i)^{-\frac{\alpha}{2}}h_is_i}_{\boldsymbol{i}_{agg}(r_0)}+\underbrace{\sum_{i\in\boldsymbol{\boldsymbol{\Psi}}_J}\sqrt{P_J\chi^J_i}(1+d_i)^{-\frac{\alpha}{2}}g_ij_i}_{\boldsymbol{j}_{agg}}+n,
\end{align}
where the BSs at distances $r_i$ from the origin send symbols $s_i\in \mathcal{M}$ that are taken from a digital amplitude phase modulated constellation $\mathcal{M}$ such as BPSK or QPSK with $\mathbb{E}(|s_i|^2)=1$. The random variable $\chi_i=\exp(x_i)$ has a log-normal distribution and models the shadowing such that $x_i\sim \mathcal{N}(\mu_{\chi},\sigma^2_{\chi})$, where $\mu_{\chi}$ and $\sigma_{\chi}$ are respectively the mean and standard deviation of $x_i$. In \eqref{eq:signal_model}, $h_i$ indicates a complex Gaussian random variable that models Rayleigh fading with $\mathbb{E}(|h_i|^2)=1$. The variables $r_0, s_0, \chi_0$ and $h_0$ are the respective parameters for the serving BS/AP with which the victim receiver communicates. All other BSs signals are therefore interfering with the serving BS signal to the victim receiver. In other words,  we assume universal frequency reuse. Without loss of generality, it is assumed that all the BSs transmit signals at power levels $P_T$ (a common assumption made in the literature \cite{PPP_Dhillon}). In \eqref{eq:signal_model}, $a_i$ is an indicator variable that indicates whether or not the $i$th BS is active at a given time instant. We assume that the interfering BSs (the ones other than the serving BS) independently transmit signals with probability $p$, also known as the activity factor or the network loading factor \cite{DhillonActivityFactor}. In other words, $p$ indicates the average fraction of the BSs that are active at any time instant once the serving BS association is completed  \cite{DhillonActivityFactor}. Therefore, $a_i=1$ with probability $p$ and is $0$ otherwise. 

In \eqref{eq:signal_model}, $\alpha$ indicates the standard power-law path loss exponent such that $\alpha>2$. It has been shown in \cite{PathLoss1}-\hspace{-0.5pt}\cite{PathLoss3} that the commonly used distance-based path loss model $r_i^{-\alpha}$ is inaccurate for smaller values of $r_i$ and that it is used only for analytical tractability. Therefore, in this paper, we use a more realistic model given by $(1+r_i)^{-\alpha}$ to model the path loss between the $i$th BS and the victim receiver. This model also avoids discontinuity at the origin \cite{PathLoss1}-\hspace{-0.5pt}\cite{PathLoss3}. In Section~\ref{sec:Results}, we show that this model results in an interesting concave-type behavior for the jammer network distribution as a function of the BS/AP density and the shadowing levels. In what follows, it is assumed that $h_i$ and $\chi_i$ are respectively independent and identically distributed. 

The jammers are located on a compact disk of radius $R_J$ centered at the origin denoted by $\mathbb{b}(0,R_J)\subset \mathbb{R}^2$. Let $N_{J_c}=\frac{N_J}{\pi R_J^2 \lambda_T}$ indicate the number of jammers per cell (or per BS). In Section~\ref{sec:Results}, we will discuss how $N_{J_c}$ affects the jamming performance at the victim receiver. The jammers attack the wireless network by sending signals $j_i\in\mathcal{M}_J$, where $\mathcal{M}_J$ indicates the signaling scheme employed by the jammers. The random variable $g_i$ is a zero-mean complex Gaussian random variable that models Rayleigh fading such that $\mathbb{E}(|g_i|^2)=1$. The random variable $\chi^J_i$ models the log-normal shadowing such that $\chi^J_i=\exp(x^J_i)$ where $x^J_i\sim \mathcal{N}(\mu_{\chi},\sigma^2_{\chi})$. The jammers send signals at a constant power level $P_J$ in order to attack the wireless network. In \eqref{eq:signal_model}, $n$ indicates the zero-mean complex additive white Gaussian noise (AWGN) as seen at the victim receiver. Define the reference signal-to-noise-ratio as $\mathtt{SNR}=\frac{P_T}{\sigma^2}$. This is termed the transmit $\mathtt{SNR}$. Along similar lines, we define the reference jammer-to-noise ratio as $\mathtt{JNR}=\frac{P_J}{\sigma^2}$. 

We assume that the shadowing is constant over the time of interest and hence the serving BS is selected based on the average received signal strength. In other words, shadowing impacts the BS selection but fading does not. Under such conditions, the displacement theorem \cite{KeelerDisplacement}, \cite{DhillonDisplacement} states that the overall effect of shadowing can be taken into consideration as a perturbation in the locations of the BSs (recall that the BSs are distributed according to a PPP) if $\mathbb{E}(\chi_i^{2/\alpha})<\infty$. When this condition holds true, without loss of generality, a new equivalent PPP with density $\lambda_T\mathbb{E}(\chi_i^{2/\alpha})$ can be defined to model the BS locations \cite{DhillonDisplacement}. Now, the strongest BS association policy in the original PPP maps to being equivalent to the nearest BS policy association in the transformed PPP without shadowing. Therefore, $r_0, s_0$ and $h_0$  in \eqref{eq:signal_model} will now represent the parameters of the signal received from the closest BS in the transformed PPP. In what follows, for the ease of notation, we denote $\lambda_T\mathbb{E}(\chi_i^{2/\alpha})$ as $\lambda_T$. Note that the condition $\mathbb{E}(\chi_i^{2/\alpha})<\infty$ is satisfied in most practical settings that usually consider log-normal shadowing with finite mean and standard deviation \cite{KeelerDisplacement}, \cite{DhillonDisplacement}. 

Since the BSs are modeled according to a PPP, $r_0$ is a random variable with probability density function (pdf) equal to $f_{r_0}(\eta)=2\pi\lambda_T\eta\exp(-\pi\lambda_T\eta^2)$ \cite{DiRenzo_TCOM}. The interference caused by the BSs besides the serving BS i.e., $\boldsymbol{\boldsymbol{\Psi}}\backslash\{0\}$, is collectively denoted by $\boldsymbol{i}_{agg}(r_0)$. The interference caused by jammers is collectively denoted by $\boldsymbol{j}_{agg}$. The jammers can transmit either additive white Gaussian noise (AWGN) or any standard modulation scheme in order to attack the receiver \cite{ModulationJamming_Journal}. The performance when the jammers use different types of jamming signals will be discussed in detail in Section~\ref{sec:Results}. A list of notations used is shown in Table~\ref{tab:notation}.

\begin{table}
\centering
\caption{Notations used}
\renewcommand{\arraystretch}{1.3}
{\fontsize{10}{10}\selectfont
\begin{tabular}{|c|c|}  \hline
\textbf{Notation} & \textbf{Definition} \\ \hline \hline
$\boldsymbol{\boldsymbol{\Psi}}$, $\lambda_T$ & PPP network of BSs/APs; density of BSs/APs  \\ \hline
$\boldsymbol{\boldsymbol{\Psi}}_J$, $N_J$  & BPP network of jammers; number of jammers \\ \hline
$N_{J_c}$ & Number of jammers per cell (per BS)\\ \hline
$P_T, P_J$ & Transmit power of BSs, jammers \\ \hline
$r_0,h_0,s_0$ & Distance, channel and symbols of the closest BS/AP \\ \hline
$r_i,h_i,s_i$ & Distance, channel and symbols of the $i$th BS/AP \\ \hline
$a_i,p$ & Indicator variable; activity factor for interfering BSs \\ \hline
$d_i,g_i,\chi^J_i,j_i$ & Distance, channel, shadowing and symbols of the $i$th jammer \\ \hline
$\mu_{\chi},\sigma_{\chi}$ & shadowing parameters \\ \hline
$\alpha$ & Power-law path loss exponent \\ \hline
\end{tabular}
}
\label{tab:notation}
\end{table}

\section{Outage probability of the victim receiver}\label{sec:CoverageProbability}
We first analyze the network performance from an outage probability perspective. When the interference is treated as noise, the signal-to-interference-plus-noise ratio $(\mathtt{SINR})$ is the most appropriate metric that captures the link quality from the serving BS to the victim receiver. However, since we are interested in interference-limited scenarios, we consider the signal-to-interference ratio $(\mathtt{SIR})$ as the metric for the outage probability analysis (all the analysis presented below can be extended in a straightforward manner to consider $\mathtt{SINR}$). For the system model defined in \eqref{eq:signal_model}, the $\mathtt{SIR}$
is given by 
\begin{align}\label{eq:sir_typical_receiver}
\mathtt{SIR}=\frac{P_T|h_0|^2 (1+r_0)^{-\alpha}}{\sum_{i \in \boldsymbol{\boldsymbol{\Psi}}\backslash\{0\}}a_iP_T|h_i|^2(1+r_i)^{-\alpha}+\sum_{i\in \boldsymbol{\boldsymbol{\Psi}}_J}P_J\chi^J_i|g_i|^2(1+d_i)^{-\alpha}}.
\end{align}
In order to effectively attack the wireless network, the jammer network intends to maximize the outage probability denoted by $P_o(\mathtt{SIR}<\theta)$ at the victim receiver. Here $\theta$ indicates a $\mathtt{SIR}$ threshold necessary for successful communication. We next derive the outage probability of the victim receiver whose $\mathtt{SIR}$ is given by \eqref{eq:sir_typical_receiver}. 
\begin{theorem}
The outage probability of a victim receiver, assumed to be at the origin, in the presence of BSs distributed according to a PPP and jammers distributed as a BPP in $\mathbb{b}(0,R_J)$ is given by
\begin{align}\label{BPP_Pc}
\mathbb{P}_{o}(\mathtt{SIR}<\theta)=&1-\int_{r_0=0}^{\infty}\exp\left[-2\pi p\lambda_T\int_{r=r_0}^{\infty}\left[1-\frac{1}{1+\theta (1+r_0)^{\alpha}(1+r)^{-\alpha}}\right]r\mathrm{d}r\right]\times\nonumber \\
&\hspace{25pt}\left[\frac{2}{R_J^2}\int_{r=0}^{R_J}\int_{\chi=0}^{\infty}\frac{1}{1+\frac{\theta (1+r_0)^{\alpha}P_J\chi}{P_T}(1+r)^{-\alpha}}\frac{r\exp\left[-\left[\frac{\log(\chi)}{\sqrt{2\sigma_{\chi}}}\right]^2\right]}{\chi\sqrt{2\sigma_{\chi}}}\mathrm{d}\chi\mathrm{d}r\right]^{N_J} \times \nonumber \\
&\hspace{25pt}2\pi\lambda_Tr_0\exp(-\lambda_T \pi r_0^2)\mathrm{d} r_0
\end{align} 
\end{theorem}
\emph{Proof:} See Appendix A. 

The outage probability in \eqref{BPP_Pc} can be simplified by noticing that
\begin{align}\label{simplify_1}
\int_{r=r_0}^{\infty}\left[1-\frac{1}{1+\theta (1+r_0)^{\alpha}(1+r)^{-\alpha}}\right]r\mathrm{d}r&=\int_{q=1+r_0}^{\infty}\frac{q-1}{1+\frac{q^{\alpha}}{\theta (1+r_0)^{\alpha}}}\mathrm{d}q, 
\end{align}
and by using the following integral
\begin{align}
\int \frac{x-1}{1+\frac{x^\alpha}{a}}\mathrm{d}x={}_2F_1\left[1,\frac{2}{\alpha};1+\frac{2}{\alpha};\frac{-x^\alpha}{a}\right]\frac{x^2}{2}
-{}_2F_1\left[1,\frac{1}{\alpha};1+\frac{1}{\alpha};\frac{-x^\alpha}{a}\right]\frac{x}{a},
\end{align}
where ${}_pF_q(a_1,\ldots,a_p;b_1,\ldots,b_q;x)$ is the generalized Hypergeometric function given by \cite{Integrals_Book}
\begin{align}\label{eq:hyp_geom_series}
{}_pF_q(a_1,\ldots,a_p;b_1,\ldots,b_q;x)=\sum_{n=0}^{\infty}\frac{(a_1)_n(a_2)_n\ldots(a_p)_n}{(b_1)_n(b_2)_n\ldots(b_q)_n}\frac{x^n}{n!},
\end{align}
where $(a_i)_n$ is the Pochhammer symbol given by $\Gamma(a_i+n)/\Gamma(a_i)$ and $\Gamma(.)$ is the Gamma function. Along similar lines, we can simplify
\begin{align}
\int_{r=0}^{R_J}\frac{1}{1+\frac{\theta (1+r_0)^{\alpha}P_J\chi}{P_T}(1+r)^{-\alpha}}r\mathrm{d}r
\end{align}
by using the following integral
\begin{align}\label{eq:BPP_Approximation}
\int \frac{(x-1)x^\alpha}{a+x^{\alpha}}=\frac{x^2}{2}-{}_2F_1\left[1,\frac{2}{\alpha};1+\frac{2}{\alpha};\frac{-x^\alpha}{a}\right]\frac{x^2}{2}-x+{}_2F_1\left[1,\frac{1}{\alpha};1+\frac{1}{\alpha};\frac{-x^\alpha}{a}\right]x.
\end{align}
Finally, by using \eqref{simplify_1}-\hspace{-0.5pt}\eqref{eq:BPP_Approximation}, the integrals in outage probability in \eqref{BPP_Pc} can be simplified. 

Note that the outage probability $\mathbb{P}_{o}(\mathtt{SIR}<\theta)$ is dependent on $\theta,N_J,P_T,P_J,\lambda_T,p$ and $\sigma_{\chi}$. The effect of each of these parameters on the outages caused by the jammer network at the victim receiver will be studied in detail in Section~\ref{sec:Results}. 

\begin{corollary}
The minimum number of jammers $N_J^*$, distributed according to a BPP, necessary to disrupt the victim network i.e., to achieve a required outage probability denoted by $\mathbb{P}_{o,th}$ at the victim receiver is given by
\begin{align}\label{eq:NJ_Optimize}
N_J^*=\arg \min _{N_J}\left|\mathbb{P}_{o}(\theta,N_J,P_T,P_J,\lambda_T,p,\sigma_{\chi})-\mathbb{P}_{o,th}\right|,
\end{align}
where $\mathbb{P}_{o}(\theta,N_J,P_T,P_J,\lambda_T,p,\sigma_{\chi})$ is given by \eqref{BPP_Pc}. 
\end{corollary}

\textbf{Gaussian-Hermite Quadrature Approximation:} Notice that evaluating \eqref{BPP_Pc} with respect to the log-normal distribution of $\chi$ is complicated. Furthermore, this procedure is computationally intensive when we intend to evaluate \eqref{eq:NJ_Optimize} in order to find the optimal number of jammers to attack a network. Hence, we propose to use the Gaussian-Hermite quadrature (GHQ) approximation for evaluating 
\begin{align}
\int_{r=0}^{R_J}\int_{\chi=0}^{\infty}\frac{1}{1+\frac{\theta (1+r_0)^{\alpha}P_J\chi}{P_T}(1+r)^{-\alpha}}\frac{r\exp\left[-\left[\frac{\log(\chi)}{\sqrt{2\sigma_{\chi}}}\right]^2\right]}{\chi\sqrt{2\sigma_{\chi}}}\mathrm{d}\chi\mathrm{d}r
\end{align}
which is the second term in \eqref{BPP_Pc}. Specifically, the GHQ approximation states that 
\begin{align}\label{eq:GHQ_Define}
\int_{-\infty}^{\infty}\exp(-x^2)f(x)\mathrm{d}x\approx \sum_{n=1}^Nw_nf(x_n),
\end{align}
where $w_n$ and $x_n$ are the weights and roots of the Hermite polynomial \cite{Abramovitz}. This approximation is known to be exact if $f(x)$ is a polynomial of degree $2N-1$. For other general functions, this approximation sum is known to converge to the integral when $N\rightarrow\infty$ \cite{GHQ_Approx_N_Infinity}. 

For the case of a log-normal distribution, using the GHQ approximation for a function $f(\chi)$, we have by an appropriate change of variables
\begin{align}\label{eq:GHQ_Approx}
\hspace{-8pt}\int_{0}^{\infty}\hspace{-8pt}f(\chi)\frac{\exp\left[-\left[\frac{\log(\chi)}{\sqrt{2\sigma_{\chi}}}\right]^2\right]}{\chi\sqrt{2\sigma_{\chi}}}\mathrm{d}\chi&=\frac{1}{\sqrt{\pi}}\int_{-\infty}^{\infty}\hspace{-8pt}f(\sqrt{2\sigma_{\chi}}y)\exp(-y^2)\mathrm{d}y
\approx \frac{1}{\sqrt{\pi}}\sum_{n=1}^Nw_nf(\sqrt{2\sigma_{\chi}}x_n).
\end{align}
Therefore, by using \eqref{eq:BPP_Approximation} and \eqref{eq:GHQ_Approx}, the double integral in the evaluation of the BPP interference at the victim receiver shown in \eqref{BPP_Pc} is now changed to a summation over $N$ terms each defined by Hypergeometric functions. Of course, the accuracy of the approximation depends on the number of terms $N$ used in \eqref{eq:GHQ_Approx}. In Section~\ref{sec:Results}, we show that $N=15$ terms are sufficient to obtain close approximations for the outage probability. 

\section{Error Probability}\label{sec:ErrorProbability}
As mentioned earlier, the outage probability analysis considers the power levels of the various signals involved to estimate $\mathtt{SIR}$ and outage probability. But this analysis alone cannot capture the overall jamming impact at the victim receiver. For instance, in our recent works \cite{ModulationJamming_Journal}, \cite{FadingJamming_Journal}, we showed that considering the victim and the jammer signaling schemes explicitly in the error probability analysis and not treating them as AWGN gives interesting insights into the jamming behavior. Therefore, in this section we evaluate the error probability of the victim receiver and investigate whether or not the results in \cite{ModulationJamming_Journal}, \cite{FadingJamming_Journal} related to jamming against a point-to-point link hold true for the case of wireless networks. In the analysis that follows, we assume that the victim receiver has perfect channel state information of the serving BS i.e., the BS which is closest to it at a distance $r_0$.

From a stochastic geometry perspective, the error probability analysis in a non-jamming scenario was studied recently in \cite{DiRenzo}-\hspace{-0.5pt}\cite{DiRenzo_TCOM} by explicitly considering the signal levels and not the power levels of the various signals involved. The error probability analysis in this paper is not a straightforward extension of the analysis in \cite{DiRenzo}-\hspace{-0.5pt}\cite{DiRenzo_TCOM} because the system model considered in this paper differs in the following aspects: a) a BPP model for the jammers that attack the victim receiver is considered, b) the effects of shadowing are introduced, and c) a realistic path loss model given by $(1+r_i)^{-\alpha}$ is considered which is different from the path loss models used in \cite{DiRenzo}-\hspace{-0.5pt}\cite{DiRenzo_TCOM}. In addition to obtaining analytical expressions for the error probability of the victim receiver, we discuss in detail the impact on the error probability of the victim receiver as a function of the various jamming signaling schemes (i.e., the various digital amplitude-phase modulation schemes) that may be used by the jammers. 

The maximum likelihood-based demodulator for decoding the symbol $s_0$ at the victim receiver when the received signal is given by \eqref{eq:signal_model} is
\begin{align}\label{eq:ml_metric}
\hat{s}_0=\arg \min_{\tilde{s}_0\in \mathcal{M}}\Big\{\Lambda(\tilde{s}_0)=|y-\sqrt{P_T}h_0(1+r_0)^{-\frac{\alpha}{2}}\tilde{s}_0|^2\Big\}.
\end{align}
By ignoring the constant energy terms, this expression can be further simplified as
\begin{align}
\Lambda(\tilde{s}_0)\propto P_T|\Delta_{s_0,\tilde{s}_0}|^2|h_0|^2(1+r_0)^{-\alpha}+2(1+r_0)^{-\frac{\alpha}{2}}\sqrt{P_T}\mathfrak{R}\left(v(r_0)h_0^*\Delta^*_{s_0,\tilde{s}_0}\right),
\end{align}
where $\Delta_{s_0,\tilde{s}_0}=s_0-\tilde{s}_0$, $v(r_0)=i_{agg}(r_0)+j_{agg}+n$ indicates the total aggregate interference, $\mathfrak{R}(x)$ indicates the real part of the variable $x$, and $x^*$ indicates the complex conjugate of $x$.

It is clear that in order to analyze the error probability of the victim receiver, $v(r_0)$ has to be characterized. This entails characterizing the statistics of $i_{agg}(r_0)$ and $j_{agg}$. In \cite{MoeWin}, the interference generated by a Poisson network model i.e., $i_{agg}(r_0)$ was shown to be equivalent in distribution to an alpha-stable distribution. This equivalence was exploited in \cite{DiRenzo} and \cite{ElSawy} to explicitly characterize the $\mathtt{SINR}$ in a non-jamming scenario (in terms of the signal levels as opposed to power levels that are commonly used to analyze outage probability, see \cite{DiRenzo}, \cite{ElSawy} for more details) to evaluate the error probability at a receiver. By using the signal-based formulation as opposed to power level-based formulation, the explicit dependency of the error probability on the modulation schemes employed by the receiver were addressed in \cite{DiRenzo} and \cite{ElSawy}. However, as noted in \cite{Haenggi_BPP} and references therein, there are no closed form approximations for the interference
originating from a Binomial field which is the model used for the jammer network distribution in this paper. Hence, alternate techniques are necessary to analyze the error probability of the victim receiver considered in this paper. Specifically, we use the nearest neighbor approximation (corresponding to the modulation scheme $\mathcal{M}$) method \cite{DiRenzo_TCOM} that provides exact expressions for binary modulations and approximations for higher order modulations. 

The following steps are used to obtain the overall average symbol error probability ($\mathtt{ASEP}$):
\begin{enumerate}
\item The pairwise error probability ($\mathtt{PEP}$) conditioned on $r_0$, $h_0$ and the specific realization of the jammer network $\boldsymbol{\Psi}_J$ is first expressed as a function of the aggregate interference $v(r_0)$. 
\item Then using the Gil-Pelaez transform \cite{DiRenzo_Gil}, we obtain the cumulative distribution function for the aggregate interference $v(r_0)$ which is used to obtain the average pairwise error probability $(\mathtt{APEP})$ by averaging over $r_0$, $h_0$ and the jammer network statistics i.e., the BPP model.
\item Finally, $\mathtt{APEP}$ is then used to compute the $\mathtt{ASEP}$ using the nearest neighbor (NN) approximation corresponding to $\mathcal{M}$ \cite{DiRenzo_TCOM}. 
\end{enumerate}

\subsection{$\mathtt{PEP}$ derivation}\label{sec:pep_derive}
Let $\mathbb{P}(s_0\rightarrow \hat{s}_0|h_0,r_0,\boldsymbol{\boldsymbol{\Psi}}_J)$ indicate the probability with which an error is made in detecting the actual symbol $s_0$ as $\hat{s}_0$. Note that this happens when the likelihood metric is maximized or in other words \eqref{eq:ml_metric} is minimized at a symbol $\hat{s}_0$ which is different from $s_0$. Mathematically, this can be represented as follows;
\begin{align}
&\hspace{-10pt}\mathbb{P}(s_0\rightarrow \hat{s}_0|h_0,r_0,\boldsymbol{\boldsymbol{\Psi}}_J)=\mathbb{P}(\Lambda(\tilde{s}_0=\hat{s}_0)<\Lambda(\tilde{s}_0=s_0))=\mathbb{P}(\Lambda(\tilde{s}_0=\hat{s}_0)<0) \nonumber \\
&\hspace{-10pt}=\mathbb{P}\Big\{\mathfrak{R}\left(v(r_0)h_0^*\Delta_{s_0,\hat{s}_0}^*\right)<-\frac{(1+r_0)^{-\frac{\alpha}{2}}\sqrt{P_T}|\Delta_{s_0,\hat{s}_0}|^2|h_0|^2}{2}\Big\} \nonumber \\
&\hspace{-10pt}\stackrel{(a)}{=}\mathbb{P}\Big\{\mathfrak{R}\left(v(r_0)\right)<-\frac{(1+r_0)^{-\frac{\alpha}{2}}\sqrt{P_T}|\Delta_{s_0,\hat{s}_0}||h_0|}{2}\Big\} = F_{v_{Re}}\left[-\frac{(1+r_0)^{-\frac{\alpha}{2}}\sqrt{P_T}|\Delta_{s_0,\hat{s}_0}||h_0|}{2}\right],
\end{align}
where $F_{v_{Re}}$ indicates the cumulative distribution function (cdf) of $v_{Re}=\mathfrak{R}(v(r_0))$. In the above equation, $(a)$ follows from the fact that conditioned on $r_0$, $|h_0|$ and the realization $\boldsymbol{\boldsymbol{\Psi}}_J$, the aggregate interference $v(r_0)$ is a circularly symmetric random variable. This is because the constellation symbols are equally probable and symmetric, $g_i$ and $h_i$ are circularly symmetric complex Gaussian random variables and $\chi^J_i$ is a real random variable. Hence, $v(r_0)$ has the same distribution as $v(r_0)\exp(-j(\theta_0+\angle{\Delta_{s_0,\hat{s}_0}^*}))$ ($\angle{x}$ indicates the phase of $x$).

Let $\Phi_x$ indicate the characteristic function of a random variable $x$. By using the Gil-Pelaez transform \cite{DiRenzo_TCOM} to express the cdf $F_{v_{Re}}(x)$ as a function of the characteristic function of the aggregate interference and the facts that (i) $v$ is a circularly symmetric random variable which implies that $\Phi_v(|\omega|;r_0)=\mathfrak{R}(\Phi_v(|\omega|;r_0))$, $\Phi_v(|\omega|;r_0)=\Phi_{v_{Re}}(|\omega|;r_0)$ \cite{ElSawy}, and (ii) $\Phi_{v_{Re}}$ is a real function, 
we have 
\begin{align}\label{eq:GilPelaez}
&F_{v_{Re}}(v)=\frac{1}{2}+\frac{1}{\pi}\int_{0}^{\infty}\frac{\sin(|\omega| v)\Phi_{v_{Re}}(|\omega|;r_0)}{|\omega|}\mathrm{d}|\omega|, \nonumber \\
&\mathtt{PEP}(\Delta_{s_0,\hat{s}_0};|h_0|,r_0,\boldsymbol{\boldsymbol{\Psi}}_J)=\frac{1}{2}-\frac{1}{\pi}\int_{0}^{\infty}\frac{\sin\left[\frac{|\omega|(1+r_0)^{-\frac{\alpha}{2}}\sqrt{P_T}|\Delta_{s_0,\hat{s}_0}||h_0|}{2}\right]\Phi_v(|\omega|;r_0)}{|\omega|}\mathrm{d}|\omega|.
\end{align} 
Please see \cite[Appendix II]{DiRenzo_TCOM} for more details on the derivation of \eqref{eq:GilPelaez}.

The next step is to evaluate $\Phi_v(|\omega|;r_0)$. Since $i_{agg}(r_0)$, $j_{agg}$ and $n$ are independent of each other we have
\begin{align}
\Phi_v(|\omega|;r_0)=\Phi_{i_{agg}}(|\omega|;r_0)\Phi_{j_{agg}}(|\omega|)\Phi_n(|\omega|).
\end{align}
Therefore the $\mathtt{APEP}$ is given by
\begin{align}\label{eq:APEP}
\mathtt{APEP}&=\mathbb{E}_{|h_0|,r_0,\boldsymbol{\boldsymbol{\Psi}}_J}\left[\mathtt{PEP}(\Delta_{s_0,\hat{s}_0};|h_0|,r_0,\boldsymbol{\boldsymbol{\Psi}}_J)\right] \nonumber \\
&\hspace{-0pt}=\frac{1}{2}-\frac{1}{\pi}\int_{0}^{\infty}\mathbb{E}_{r_0}\Bigg\{\mathbb{E}_{|h_0|}\Big\{\frac{\sin\left[\frac{|\omega|(1+r_0)^{-\frac{\alpha}{2}}\sqrt{P_T}|\Delta_{s_0,\hat{s}_0}||h_0|}{2}\right]}{|\omega|}\Big\}\Phi_{i_{agg}}(|\omega|;r_0)\Bigg\}\times\nonumber \\
&\hspace{280pt}\mathbb{E}_{\boldsymbol{\boldsymbol{\Psi}}_J}\left[\Phi_{j_{agg}}(|\omega|)\right]\Phi_n(|\omega|)\mathrm{d}|\omega| \nonumber \\
&\hspace{-0pt}\stackrel{(i)}{=}\frac{1}{2}-\frac{1}{\pi}\int_{0}^{\infty}E_{r_0}\Bigg\{\frac{\sqrt{\pi P_T}}{4(1+r_0)^{\frac{\alpha}{2}}}|\Delta_{s_0,\hat{s}_0}|\exp\left[-\frac{P_T|\Delta_{s_0,\hat{s}_0}|^2|\omega|^2}{16(1+r_0)^{\alpha}}\right]\Phi_{i_{agg}}(|\omega|;r_0)\Bigg\}\times\nonumber\\
&\hspace{250pt}\mathbb{E}_{\boldsymbol{\boldsymbol{\Psi}}_J}\left[\Phi_{j_{agg}}(|\omega|)\right]\Phi_n(|\omega|)\mathrm{d}|\omega|, 
\end{align}
where $(i)$ follows by using the fact that $h_0$ is a zero-mean complex Gaussian random variable with unit variance i.e., $|h_0|$ is a Rayleigh random variable. Thus, we have to first evaluate the characteristic functions of $i_{agg}(r_0)$ and $j_{agg}$. It is known that $\Phi_n(|\omega|)=\exp(-|\omega|^2/4)$ for a zero-mean, unit variance complex Gaussian random variable \cite{Papoulis}. It remains to evaluate $\Phi_{i_{agg}}(|\omega|;r_0)$ and $\mathbb{E}_{\boldsymbol{\boldsymbol{\Psi}}_J}\left[\Phi_{j_{agg}}(|\omega|)\right]$. 

The characteristic function $\Phi_{j_{agg}}(|\omega|)$ for a given realization of the jammer topology is first evaluated and then averaged over the BPP in order to obtain $\mathbb{E}_{\boldsymbol{\boldsymbol{\Psi}}_{j}}\left[\Phi_{j_{agg}}(|\omega|)\right]$. 
\begin{align}\label{eq:jam_intf_charac_initial}
&\Phi_{j_{agg}}(|\omega|)\stackrel{(i)}{=}\Phi_{j_{agg}}(\omega)=\mathbb{E}\left[\exp\left[j\omega\sum_{i=1}^{N_J}\sqrt{P_J\chi^J_i}g_ij_i(1+d_i)^{-\frac{\alpha}{2}}\right]\right] \nonumber \\
&=\prod_{i=1}^{N_J}\mathbb{E}\left[\exp\left[j\omega\sqrt{P_J\chi^J_i}g_ij_i(1+d_i)^{-\frac{\alpha}{2}}\right]\right]\stackrel{(ii)}{=}\prod_{i=1}^{N_J}\Phi_z\left[\omega\sqrt{P_J\chi^J_i}(1+d_i)^{-\frac{\alpha}{2}}\right] \nonumber \\
&\stackrel{(iii)}{=}\prod_{i=1}^{N_J}\Phi_{z_{Re}}\left[\omega\sqrt{P_J\chi^J_i}(1+d_i)^{-\frac{\alpha}{2}}\right]\stackrel{(iv)}{=}\prod_{i=1}^{N_J}\mathbb{E}_{\hat{z}}\left[\cos\left[|\omega|\sqrt{P_J}(1+d_i)^{-\frac{\alpha}{2}}\hat{z}\right]\right],
\end{align}
where $(i)$ follows by taking into account the fact that $\sqrt{\chi^J_i}g_ij_i$ is a circularly symmetric random variable \cite[Appendix I]{DiRenzo_TCOM}, the expectation is with respect to the statistics of $\chi^J_i$, $g_i$ and $j_i$ because $d_i$ is fixed for a given realization of the jammer network, $(ii)$ follows by defining a new variable $z=\sqrt{\chi^J_i}g_ij_i$, $(iii)$ and $(iv)$ hold because $z$ is a circularly symmetric random variable \cite[Appendix I]{DiRenzo_TCOM} and by a change of variable $\hat{z}=z_{Re}\triangleq\mathfrak{R}(z)$. Since $\chi^J_i$ is an independent and identically distributed (i.i.d.) random variable, in what follows we denote it by $\chi$ for the ease of notation. 

\begin{theorem}\label{theo:jammer_charac_func}
The characteristic function of the jammer interference as seen at the victim receiver i.e.,  $\mathbb{E}_{\boldsymbol{\boldsymbol{\Psi}}_J}\left[\Phi_{j_{agg}}(|\omega|)\right]$ is given by
{\small 
\begin{align}\label{eq:jammer_charac_func}
\mathbb{E}_{\boldsymbol{\boldsymbol{\Psi}}_J}\left[\Phi_{j_{agg}}(|\omega|)\right]&=\Bigg(\frac{1}{|\mathcal{M}_J|}\sum_{j_i\in \mathcal{M}_J}\mathbb{E}_{\chi}\Bigg\{\nonumber \\
&\hspace{-60pt}\frac{1}{R_J^2}\left[{}_1F_1\left[\frac{-2}{\alpha};1-\frac{2}{\alpha};-\frac{(|\omega|\sqrt{P_J\chi}|j_i|)^2(1+R_J)^{-\alpha}}{4}\right](1+R_J)^2 
-{}_1F_1\left[\frac{-2}{\alpha};1-\frac{2}{\alpha};-\frac{(|\omega|\sqrt{P_J\chi}|j_i|)^2}{4}\right]\right] \nonumber \\
&\hspace{-80pt}-\frac{2}{R_J^2}\left[{}_1F_1\left[\frac{-1}{\alpha};1-\frac{1}{\alpha};-\frac{(|\omega|\sqrt{P_J\chi}|j_i|)^2(1+R_J)^{-\alpha}}{4}\right](1+R_J)- 
{}_1F_1\left[\frac{-1}{\alpha};1-\frac{1}{\alpha};-\frac{(|\omega|\sqrt{P_J\chi}|j_i|)^2}{4}\right] \right]\Bigg\}\Bigg)^{N_J}\hspace{-10pt}.
\end{align}
}
\end{theorem}
\emph{Proof}: See Appendix B. 

In Appendix~B we show that the error probability of the victim receiver depends on the jammer's signaling scheme via the term $\mathbb{E}(\hat{z}^{2q})$ (which was encountered in \eqref{phi_J_inter} in the derivation of the $\mathbb{E}_{\boldsymbol{\boldsymbol{\Psi}}_J}\left[\Phi_{j_{agg}}(|\omega|)\right]$) given by
\begin{align}\label{eq:z_moments}
\mathbb{E}(\hat{z}^{2q})=\frac{\Gamma(q+\frac{1}{2})}{\sqrt{\pi}}E(\chi^q)\mathbb{E}\left[|j_i|^{2q}\right].
\end{align}
Here $q$ is a non-negative integer and $\Gamma(x)$ is the gamma function. 
\begin{remark}
Notice that any constant modulus signaling scheme will have the same value for $\mathbb{E}(\hat{z}^{2q})$. This indicates that irrespective of the constant modulus-signaling scheme used by the jammer, the error probability at the victim receiver will remain the same for the system model given in \eqref{eq:signal_model}. This behavior is due to the fact that the jammer is not aware of the channel $g_i$ between itself and the victim receiver and hence cannot compensate for the random rotations introduced by $g_i$. Therefore, the results in \cite{ModulationJamming_Journal} and \cite{FadingJamming_Journal} which indicate that modulation-based jamming is optimal, cannot be reproduced in this case. This behavior will be explained in detail via numerical results in Section~\ref{sec:Results}. 
\end{remark}

\begin{corollary}\label{AWGN_Corollary}
The $\mathtt{APEP}$ of the victim receiver when the jammer network uses a zero-mean, unit variance AWGN jamming signal is given by replacing 
$\mathbb{E}(|j_i|^{2q})$ in \eqref{eq:z_moments} with $\frac{2^q\Gamma(q+\frac{1}{2})}{\sqrt{\pi}}$.
\end{corollary}
\emph{Proof}: From \cite[Eq. 16]{Moments_Gaussian}, it is known that if $X$ is a Gaussian random variable with mean $\nu$ and variance $\sigma^2$, we have
\begin{align}\label{identity_2_pre}
\mathbb{E}\left[(X-\nu)^{2q}\right]=\sigma^{2q}\frac{2^{q}\Gamma(q+\frac{1}{2})}{\sqrt{\pi}}. 
\end{align}
The proof is straightforward using this result. 
\begin{remark}
Corollary~\ref{AWGN_Corollary} states that the theoretical expression for AWGN jamming can be obtained by replacing $\mathbb{E}(|j_i|^{2q})$ in \eqref{eq:z_moments} with \eqref{identity_2_pre} i.e., instead of averaging over the various modulation symbols that the jammer may use, the averaging is performed over the Gaussian distribution. While it is not easy to explain the AWGN jamming signal performance by \eqref{eq:jammer_charac_func} alone, based on the results from our earlier works in \cite{ModulationJamming_Journal}, \cite{FadingJamming_Journal} it is expected that when the jammer is not aware of the channel $g_i$, then the jamming performance of any modulation-based jamming signal and AWGN jamming signal will be  the same. However, when the jammer can compensate for the effects of the fading channel, then the error probability at the victim receiver can be significantly increased by using specific modulation schemes. We will discuss this behavior via numerical results in Section~\ref{sec:Results}. 
\end{remark}

\begin{corollary}\label{corr:poiss_char_func}
The characteristic function of the aggregate interference from the BSs other than the serving BS is given by
\begin{align}\label{eq:poiss_char_func}
\Phi_{i_{agg}}(\omega;r_0)&=\exp\left[\pi p\lambda_Tr_0^2-T_1+T_2\right],
\end{align}
where $T_1$ is given by
\begin{align}
T_1&=\frac{1}{|\mathcal{M}|}\sum_{s_i\in \mathcal{M}}\pi p\lambda_T(1+r_0)^2{}_1F_1\left[\frac{-2}{\alpha};1-\frac{2}{\alpha};-\frac{(|\omega|\sqrt{P_T}|s_i|)^2(1+r_0)^{-\alpha}}{4}\right]. \nonumber 
\end{align}
and $T_2$ is given by
\begin{align}
T_2&=\frac{1}{|\mathcal{M}|}\sum_{s_i\in \mathcal{M}}2\pi p\lambda_T(1+r_0){}_1F_1\left[\frac{-1}{\alpha};1-\frac{1}{\alpha};-\frac{(|\omega|\sqrt{P_T}|s_i|)^2(1+r_0)^{-\alpha}}{4}\right]. \nonumber 
\end{align}
\end{corollary}
\emph{Proof}: See Appendix C.
Notice that the effect of the modulation scheme used by the BSs is captured explicitly in $\Phi_{i_{agg}}(\omega;r_0)$. 

\begin{remark} Since the path loss model used in this paper is different from the ones used in \cite{DiRenzo}-\hspace{-0.5pt}\cite{DiRenzo_TCOM}, the characteristic function $\Phi_{i_{agg}}(|\omega|;r_0)$  of the interference from the BSs other than the serving BS is also different from the ones obtained in \cite{DiRenzo}-\hspace{-0.5pt}\cite{DiRenzo_TCOM}. 
\end{remark}

Finally, by using $\Phi_{i_{agg}}(|\omega|;r_0)$ and $\mathbb{E}_{\boldsymbol{\boldsymbol{\Psi}}_J}\left[\Phi_{j_{agg}}(|\omega|)\right]$, the overall $\mathtt{APEP}$ in \eqref{eq:APEP} can be evaluated irrespective of the signaling schemes used by the BSs and the jammers. 

\subsection{Gaussian-Hermite quadrature approximation}
In order to evaluate the jammer characteristic function $\mathbb{E}_{\boldsymbol{\boldsymbol{\Psi}}_J}\left[\Phi_{j_{agg}}(|\omega|)\right]$, it is necessary to evaluate functions of the form $\mathbb{E}_{\chi}\left[{}_pF_q\left[a_1,\ldots,a_p;b_1,\ldots,b_q;f\right]\right]$. Since it is computationally intensive to evaluate the integrals of hypergeometric functions, we again use the Gaussian-Hermite quadrature approximation. 
\begin{lemma}
By using the Gaussian-Hermite quadrature approximation, we have
\begin{align}\label{ghq_hypergeom}
\hspace{-10pt}\mathbb{E}_{\chi}\left[{}_pF_q\left[a_1,\ldots,a_p;b_1,\ldots,b_q;f\right]\right]\approx \frac{1}{\sqrt{\pi}}\sum_{n=1}^Nw_n{}_pF_q\left[a_1,\ldots,a_p;b_1,\ldots,b_q;f\exp(\sqrt{2}\sigma_{\chi}x_n)\right],
\end{align}
where $w_n$ and $x_n$ are the weights and roots of the Hermite polynomial \cite{Abramovitz}. 
\end{lemma}
\emph{Proof}: The proof follows by using the series expansion of the generalized Hypergeometric function ${}_pF_q\left[a_1,\ldots,a_p;b_1,\ldots,b_q;f\right]$ in \eqref{eq:hyp_geom_series} and the GHQ approximation in \eqref{eq:GHQ_Approx}. 

In Section~\ref{sec:Results}, we show that $N=10$ terms can closely approximate \eqref{ghq_hypergeom}.

\subsection{$\mathtt{ASEP}$ Evaluation}
$\mathtt{ASEP}$ can be upper bounded by using the union bound and $\mathtt{APEP}$ as follows:
\begin{align}\label{eq:asep1}
\mathtt{ASEP}\leq\frac{1}{M}\sum_{m=1}^M\sum_{i=1}^{N_{m}}\mathtt{APEP}(|\Delta_{m,i}|),
\end{align}
where $M$ is the total number of equi-probable symbols in the constellation $\mathcal{M}$, $N_{m}$ are the total number of neighbors for the $m$th symbol and $|\Delta_{m,i}|$ is the distance between the $m$th symbol and its $i$th neighbor. Notice that this expression can be evaluated fairly easily for lower order modulations such as BPSK and QPSK. However, this expression quickly becomes unwieldy for higher order modulations such as $16$-QAM. By using the nearest neighbor approximation (corresponding to $\mathcal{M}$), $\mathtt{ASEP}$ can be approximated as 
\begin{align}\label{eq:asep2}
\mathtt{ASEP}\approx\frac{1}{M}\sum_{m=1}^MN_{\Delta_{\min}}^{m}\mathtt{APEP}(|\Delta_{\min}^{m}|),
\end{align}
where $\Delta_{\min}^{m}$ is the minimum distance between the $m$th symbol and all its neighbors and $N_{\Delta_{\min}}^{m}$ is the number of such neighbors that are at a distance of 
$\Delta_{\min}^{m}$. For symmetric constellations, where $\Delta_{\min}^{m}$ is the same for all symbols, then we have $\mathtt{ASEP}\approx N_{\Delta_{\min}}^{avg}\mathtt{APEP}(|\Delta_{\min}|)$ where $N_{\Delta_{\min}}^{avg}=\frac{1}{M}\sum_{m=1}^MN_{\Delta_{\min}}^{m}$.  For instance, in the case of $16$-QAM, we have $N_{\Delta_{\min}}^{avg}=3$ and $\Delta_{\min}=2/\sqrt{10}$ (assuming unit-average energy for the modulation scheme). It is important to observe that this method gives exact error probability expression only for binary modulation schemes. We will next present several results that compare these theoretical expressions with Monte-Carlo simulations. 

\section{Results}\label{sec:Results}
In this section, numerical results are shown in order to validate the theoretical inferences presented earlier and also to shed light on the jamming impact against the wireless network in terms of the outage and error probability of the victim receiver. Unless otherwise specified, we use a BS deployment density equivalent to that of an hexagonal grid with $500$m inter site distance i.e. $\lambda_T=2/(\sqrt{3}\cdot500^2\textnormal{m}^2)$ \cite{Schloemann}. The simulation area is chosen such that an average of 100 active BSs (according to the activity factor $p$) are present in the wireless network (to avoid edge effects). The path loss exponent $\alpha$ is taken to be $3.7$, $\mu_{\chi}=0$ and $\sigma_{\chi}=6$dB. In order to account for the shadowing in the BS network, using the displacement theorem the effective BS density is taken to be $\lambda_T\exp\left[\frac{2\sigma_{\chi}^2}{\alpha^2}\right]$. The radius $R_J$ of the compact disk $\mathbb{b}(0,R_J)$ in which the jammers are distributed according to a BPP depends on $\lambda_T, N_J$ and $N_{J_c}$ as $\sqrt{\frac{N_J}{\pi\lambda_TN_{J_c}}}$. The outage and the error probabilities are functions of $P_T$, $P_J$, $\lambda_T$, $N_J$, $N_{J_c}$, $R_J$, $\sigma_{\chi}$, and $p$ and are studied next. 
\subsection{Outage Probability}
Since we consider $\mathtt{SIR}$ to study the outage probability, only $\frac{P_T}{P_J}$ matters and not their actual values. Unless otherwise mentioned, we take this to be $0$dB.
It is easy to realize that the outage probability increases (decreases) as $\frac{P_T}{P_J}$ decreases (increases). Hence, due to a lack of space, we do not discuss the impact of the jammer network as a function of  $\frac{P_T}{P_J}$.

\begin{figure}
	\centering
	\begin{minipage}{0.48\textwidth}
		\centering
		\includegraphics[width=\linewidth]{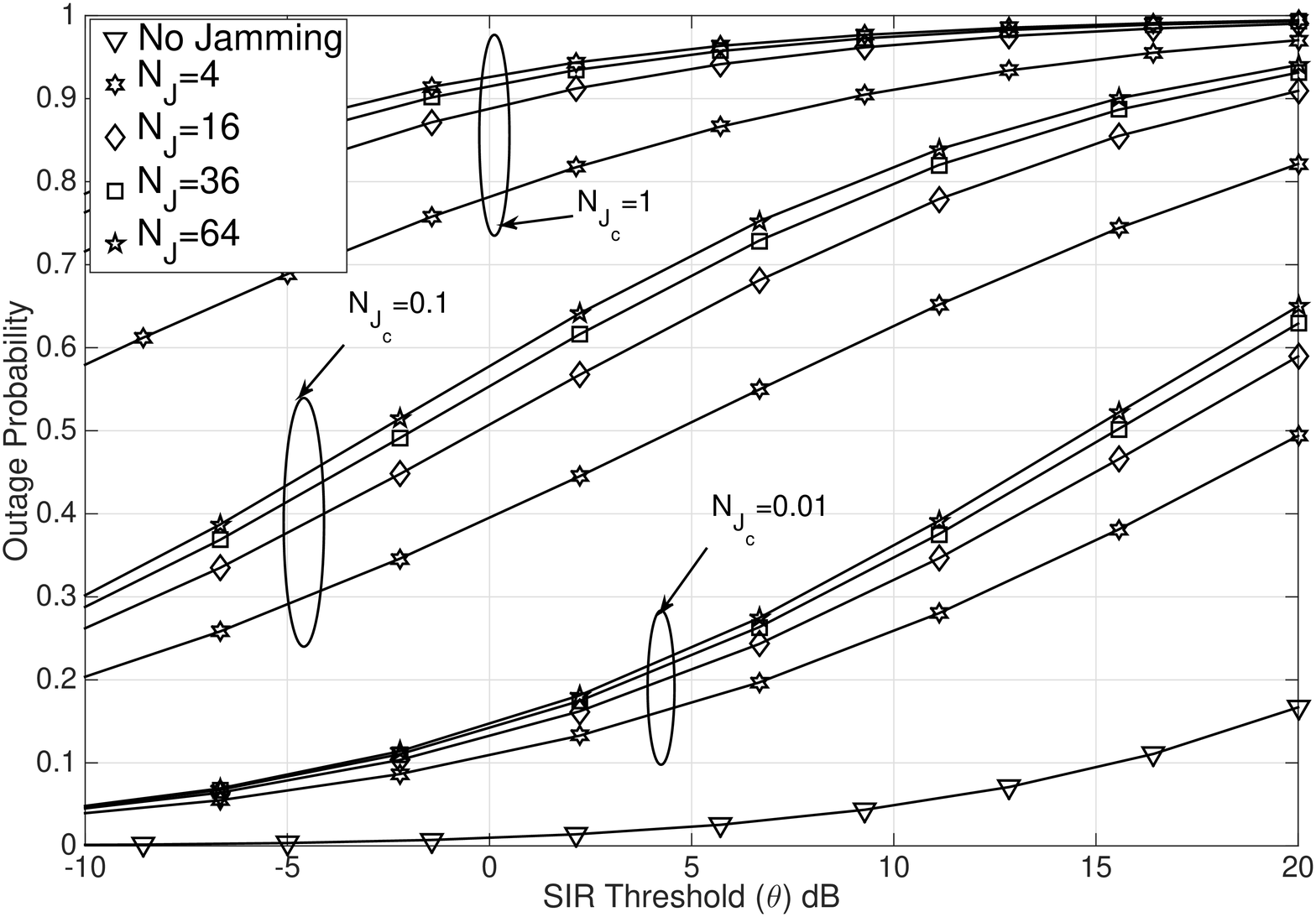}
		\vspace{-20pt}
		\caption{[Effect of $N_J$]: Outage probability of the victim receiver as a function of the number of jammers $N_J$ in the network. $p=0.01$, $P_T/P_J=0dB$. The solid lines indicate the outage probability obtained via Monte Carlo simulations and the markers indicate the theoretical outage probability evaluated using \eqref{BPP_Pc}.}
			\label{figPc}
	\end{minipage} \hfill
	\begin{minipage}{0.48\textwidth}
		\centering
		\includegraphics[width=\linewidth]{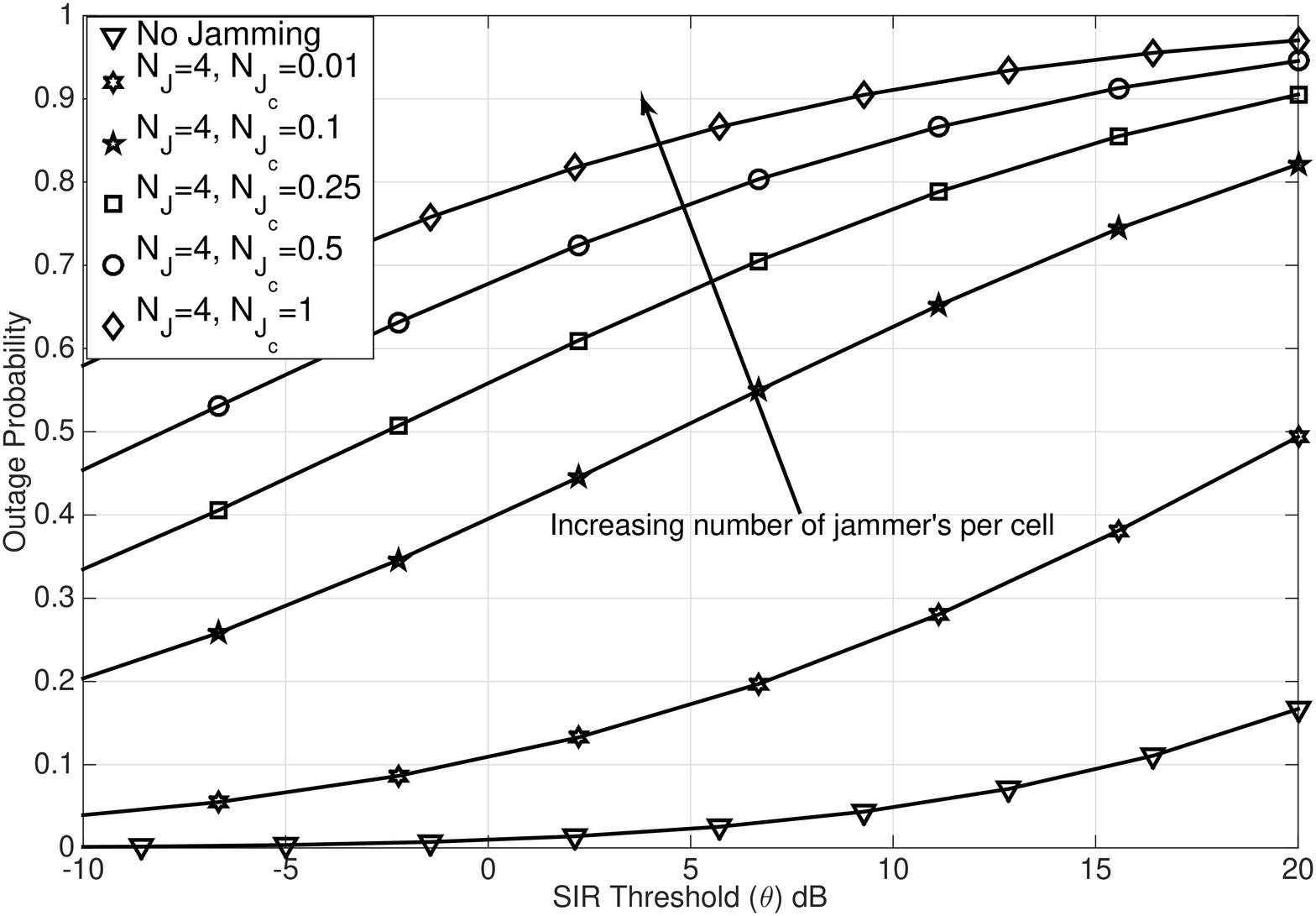}
		\vspace{-20pt}
		\caption{[Effect of $N_{J_c}$]: Outage probability of the victim receiver as a function of the number of jammers per cell (or per BS) $N_{J_c}$ in the network. $p=0.01$, $N_{J}=4$, $P_T/P_J=0dB$. The solid lines indicate the outage probability obtained via Monte Carlo simulations and the markers indicate the theoretical outage probability evaluated using \eqref{BPP_Pc}.}		
			\label{figOutage_versus_NJc}
	\end{minipage}
\end{figure}%

Fig.~\ref{figPc} shows the outage probability of the victim receiver as a function of the number of jammers in the network. Firstly see that the theoretical expression evaluated using \eqref{BPP_Pc} perfectly matches with the Monte Carlo simulations for all values of $N_J$. This validates the correctness of \eqref{BPP_Pc}. Next note that as the number of jammers increases in the network, the outage probability of the victim receiver increases due to increased interference from the jammers. However the jamming impact does not increase significantly as $N_J$ increases beyond $16$. This is because as $N_J$ increases, $R_J$ also increases in order to ensure that $N_{J_c}=1$. This allows the jammers to attack a larger region of the BS network. Since $R_J$ increases, the jamming impact at the victim receiver (which is at the origin) is limited due to path loss. In Fig.~\ref{figPc}, see that for a $1\%$ network loading factor $(p=0.01)$, the outage probability at an $\mathtt{SIR}$ threshold of $0$dB is less than $1\%$ when there are no jammers. However, with only $1$ jammer per cell, the outage probability can be increased close to $80\%$. It is interesting to see that using only a few number of jammers, the victim's link to the serving BS can be significantly degraded. 

Fig.~\ref{figOutage_versus_NJc} shows the outage probability of the victim receiver as a function of the number of jammers per cell (or per BS) in the network when $N_J$ is fixed to $4$. As $N_{J_c}$ increases, the outage probability of the victim receiver increases as expected. Again if the jammer network can ensure that there exists at least $1$ jammer per cell, then the wireless network performance can be significantly worsened. 

\begin{figure}
	\centering
	\begin{minipage}{0.48\textwidth}
		\centering
		\includegraphics[width=\linewidth]{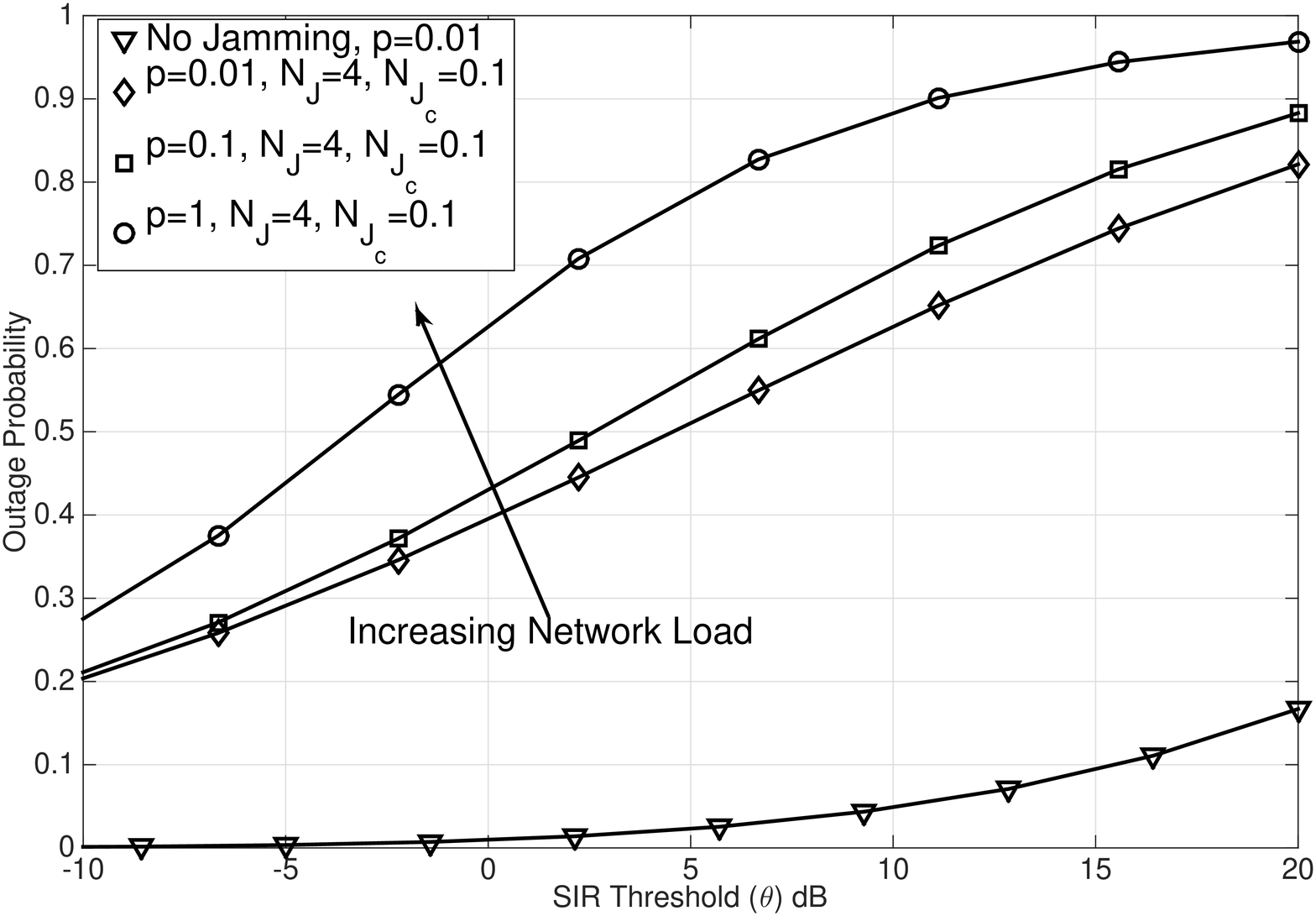}
		\vspace{-20pt}
		\caption{[Effect of $p$]: Outage probability of the victim receiver as a function of the activity factor $p$. $N_J=4$, $N_{J_c}=1$, $P_T/P_J=0dB$. The solid lines indicate the outage probability obtained using Monte Carlo simulations and the markers indicate the theoretical outage probability expression evaluated using \eqref{BPP_Pc}.}
			\label{fig:Outage_versus_p}
	\end{minipage} \hfill
	\begin{minipage}{0.48\textwidth}
		\centering
		\includegraphics[width=\linewidth]{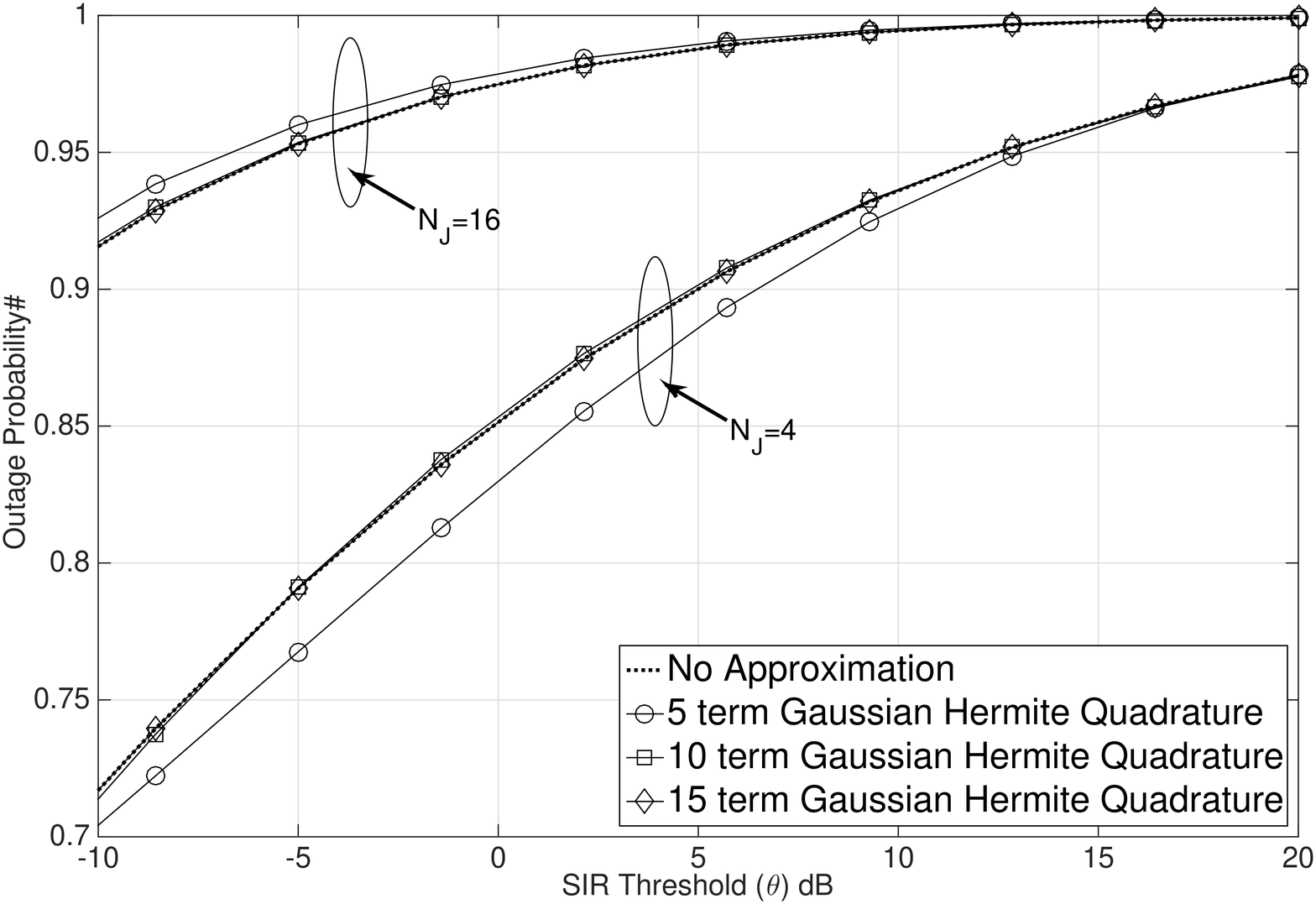}
		\vspace{-20pt}
		\caption{[GHQ Approximation]: The accuracy of the Gaussian-Hermite quadrature approximation in evaluating the outage probability as a function of the number of terms $N$ used in \eqref{eq:GHQ_Approx}. The dotted line is the outage probability evaluated using \eqref{BPP_Pc}. The marked lines indicate the outage probability evaluated using \eqref{eq:GHQ_Approx} for various values of $N$. $p=0.01$, $N_{J_c}=1$, $P_T/P_J=0dB$.}		
			\label{fig:GHQ_Approx}
	\end{minipage}
\end{figure}%

Fig.~\ref{fig:Outage_versus_p} shows the outage probability as a function of the network load/ activity factor $p$. The outage probabilities are seen to increase as the activity factor increases due to increased interference from the interfering BSs. Notice that the outage probability at loading factors of $1\%$ and $10\%$ is nearly the same. This indicates that the wireless network performance is limited by the jammer network. However, when the loading factor increases to $100\%$ i.e., all the BSs are active, then the interference significantly rises which leads to higher levels of outage probability.

Fig.~\ref{fig:GHQ_Approx} shows the accuracy of the Gaussian-Hermite quadrature approximation in evaluating the outage probability given in \eqref{BPP_Pc}. Specifically, it shows how close \eqref{eq:GHQ_Approx} is to the true outage probability when the number of terms used in the approximation is varied. It is seen that for $N=15$, the approximation is very close to the true value. Further, it is seen that this approximation holds true for different values of $N_J$. This approximation will be used next to analyze the behavior of the outage probability as a function of the various network parameters. 

We now study how the jammer network must adapt according to the various wireless network parameters in order to achieve a target outage probability. Specifically, we will study how $N_J^*$ in \eqref{eq:NJ_Optimize} varies as a function of $\lambda_T$, $\sigma_{\chi}$ and $p$ when $\mathbb{P}_{o,th}=90\%$ at a $\mathtt{SIR}$ threshold of $\theta=0$dB.
We consider two scenarios: a) when $R_J$ changes based on the network parameters, and b) when $R_J$ is fixed. For the case when $R_J$ changes based on the network parameters, we take $N_{J_c}$ to be equal to $1$. For the cases where $R_J$ is fixed, $N_{J_c}$ is dependent on $N_J$.\footnote{In a military setting where the jammers would intend to attack any device within a given region, $R_J$ is typically taken to be fixed and then the jammer network behavior is studied.}

\begin{figure}[ht]
\centering
\vspace{-10pt}
\includegraphics[width=0.55\textwidth]{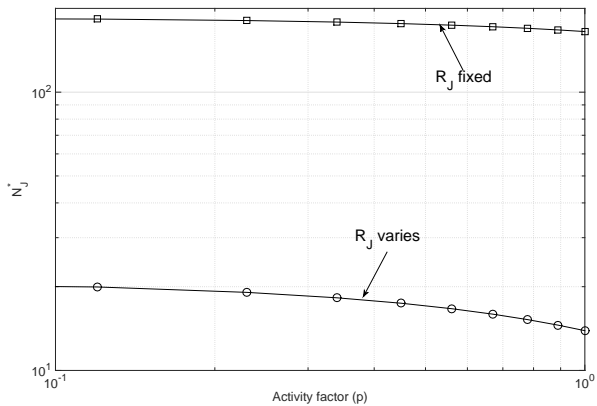}
\vspace{-10pt}
\caption{[Effect of activity factor $p$]: Number of jammers $N_J^*$ required to cause a $90\%$ probability of outage in the wireless network, as a function of the activity factor (network load) $p$. $P_T/P_J=0dB$.}
\label{fig:NJ_versus_p}
\end{figure}
Fig.~\ref{fig:NJ_versus_p} shows how the number of jammers must scale depending on the activity factor of the interfering BSs. When $R_J$ varies with the network parameters as $R_J=\sqrt{\frac{N_J}{N_{J_c}\pi\lambda_T}}$, it depends on $N_J$ (which is the optimization variable in \eqref{eq:NJ_Optimize}) and $\lambda_T=\frac{2}{\sqrt{3}\cdot500^2}\exp\left[\frac{2\sigma_{\chi}^2}{\alpha^2}\right]$. When $R_J$ is fixed, it is taken to be $\frac{10}{\sqrt{\lambda_T}}$ and in this case $N_{J_c}$ is a function of the optimization variable $N_J$. In both these cases it is seen in Fig.~\ref{fig:NJ_versus_p} that $N_J^*$ decreases with increasing values of $p$. This is because the interference from the BSs (other than the serving BS) increases as $p$ increases. Hence the number of jammers necessary to achieve $\mathbb{P}_{o,th}=90\%$ decreases. When $R_J$ varies, it is seen that $N_{J_c}$ varies inbetween $0.5$ and $0.6$ which indicates that only one jammer for every two BSs is sufficient to attack the wireless network. This value of course changes based on the value of $R_J$. 

\begin{figure}
	\centering
	\begin{minipage}{0.48\textwidth}
		\centering
		\includegraphics[width=\linewidth]{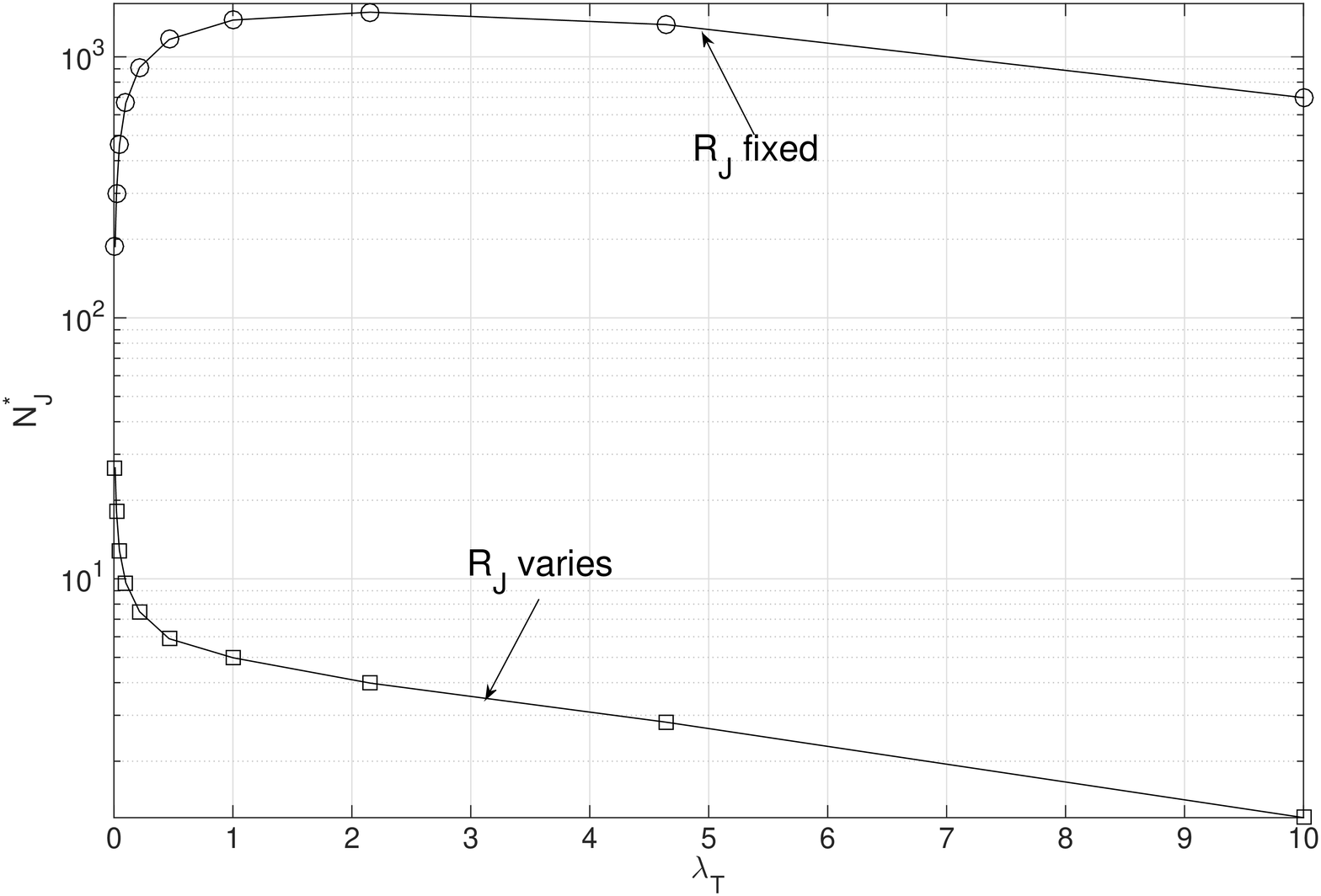}
		\vspace{-20pt}
		\caption{[Effect of $\lambda_T$]: Number of jammers $N_J^*$ required in a BPP to cause a $90\%$ probability of outage in the wireless network, as a function of $\lambda_T$, $p=0.1$.}
			\label{fig:NJ_versus_lambda}
	\end{minipage} \hfill
	\begin{minipage}{0.48\textwidth}
		\centering
		\includegraphics[width=\linewidth]{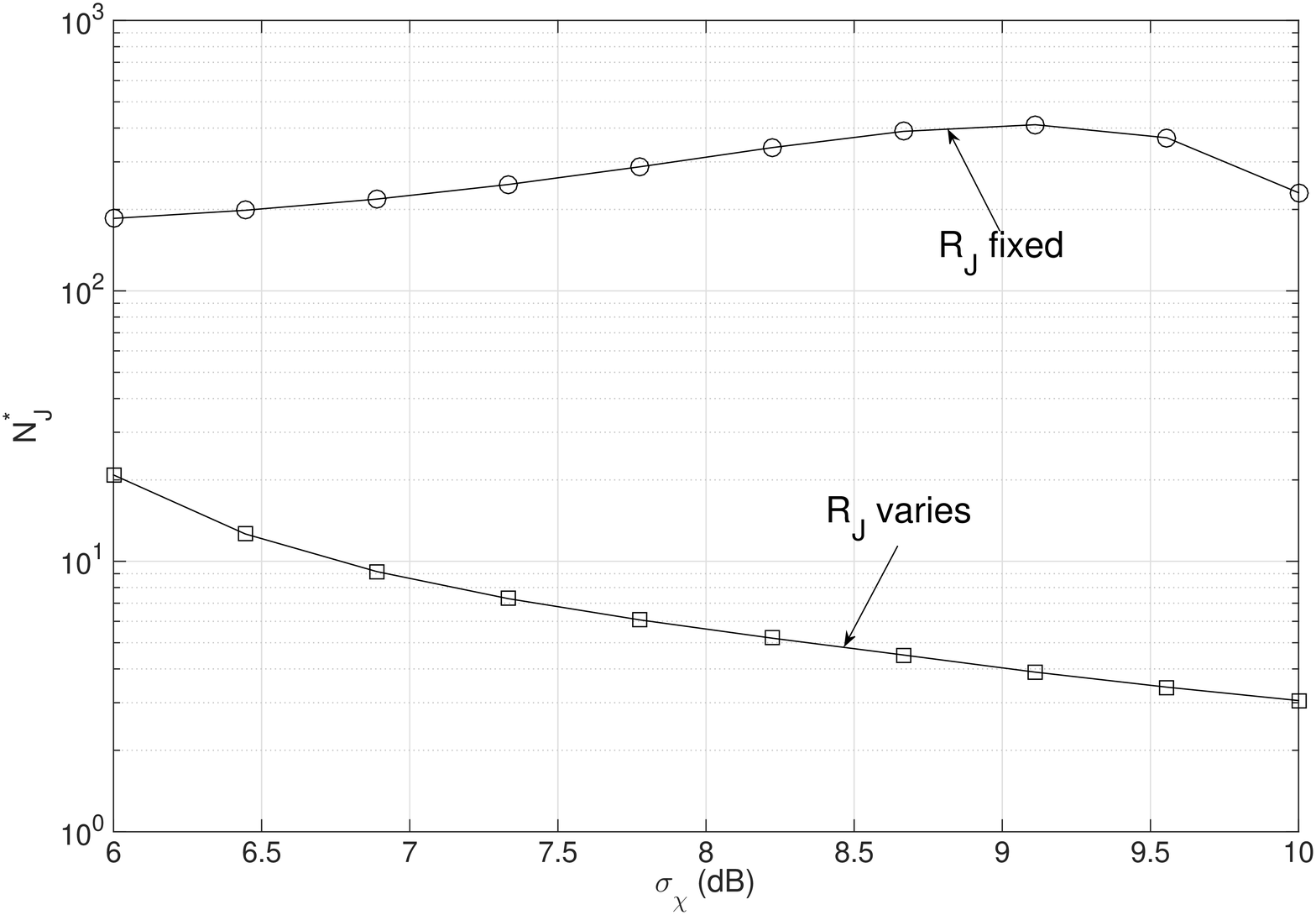}
		\vspace{-20pt}
		\caption{[Effect of Shadowing]: Number of jammers $N_J^*$ required in a BPP to cause a $90\%$ probability of outage in the wireless network, as a function of $\sigma_{\chi}$ and $p=0.01$.}		
			\label{fig:NJ_versus_Shadowing}
	\end{minipage}
\end{figure}%

Fig.~\ref{fig:NJ_versus_lambda} shows how the number of jammers must scale depending on the density of the BSs/APs in the wireless network. Before we explain the behavior of $N_J^*$ with respect to $\lambda_T$, we first discuss a non-jamming scenario result in \cite{Andrews_HetNet}, \cite{PPP_Dhillon} which is crucial to understand the behavior of $N_J^*$ in Fig.~\ref{fig:NJ_versus_lambda}. In a non-jamming scenario, the authors in \cite{Andrews_HetNet}, \cite{PPP_Dhillon} show that the outage probability remains constant irrespective of $\lambda_T$ in the interference-limited scenario. This is because the increase in interference from the BSs (besides the serving BS) is equalized by the increased signal power received from the serving BS. As $\lambda_T$ increases, the distribution of $r_0$ (distance from the serving BS) is concentrated at smaller values which indicates that the serving BS is much closer for high values of $\lambda_T$. However, this behavior is not seen in this paper due to the fact that the path loss model $(1+r_i)^{-\alpha}$ chosen in this paper is different from the path loss model $r_i^{-\alpha}$ in \cite{Andrews_HetNet}, \cite{PPP_Dhillon}. Due to the path loss model considered in this work, it can be shown that there is a limit on how much the received signal power from the serving BS can increase which thereby results in a decreasing coverage/ increasing outage probability as $\lambda_T$ increases (due to a lack of space, the formal analysis showing this behavior is skipped in this paper). Again, we believe that this is a more realistic situation in the context of wireless networks. 

When $R_J$ varies as $\sqrt{\frac{N_J}{N_{J_c}\pi\lambda_T}}$, its value decreases as $\lambda_T$ increases and hence the jammers are distributed in a smaller disk around the victim receiver. Therefore, in addition to the increased interference from the BSs (other than the serving BS), the jammer interference also increases. This compound effect indicates that a smaller number of jammers would now be needed to achieve $\mathbb{P}_{o,th}=90\%$. However, when $R_J$ is fixed, the behavior with respect to $\lambda_T$ is not obvious. An interesting concave-type behavior is seen as $\lambda_T$ increases. 

When $R_J$ is fixed at $\frac{10}{\sqrt{\min(\lambda_T)}}$ (here $\min(\lambda_T)$ is the smallest value of $\lambda_T$ considered in Fig.~\ref{fig:NJ_versus_lambda}), the jammers are always taken to be uniformly distributed within the disk $\mathbb{b}(0,\frac{10}{\sqrt{\min(\lambda_T)}})$. As mentioned above, the interference from the BSs increases as $\lambda_T$ increases. However at smaller values of $\lambda_T$, this interference is not sufficient to achieve the target outage probability $\mathbb{P}_{o,th}$. Therefore the number of jammers $N_J^*$ must increase in order to achieve $\mathbb{P}_{o,th}=90\%$. This behavior is only seen until a certain value of $\lambda_T$. Beyond this critical value of $\lambda_T$, the interference due to the BSs significantly increases at which point the number of jammers needed to achieve an outage probability of $\mathbb{P}_{o,th}=90\%$ decreases.

Fig.~\ref{fig:NJ_versus_Shadowing} shows the behavior of $N_J^*$ as a function of the standard deviation of the shadowing. As $\sigma_{\chi}$ increases, both $\lambda_T$ and the average jammer power levels received at the victim receiver increase. Therefore, based on this compound behavior and the results in Fig.~\ref{fig:NJ_versus_lambda} (which discuss the jammer impact as a function of $\lambda_T$), the behavior of $N_J^*$ in Fig.~\ref{fig:NJ_versus_Shadowing} follows. 

\textbf{Effect of Retransmissions:}
Retransmissions are used in most wireless protocols in order to improve the probability of successful communication (for delay tolerant applications). However, it is well known that retransmissions increase the interference in the network especially when the transmissions are uncoordinated. We next discuss the jamming impact against a wireless network that uses retransmissions. 

Let $D$ represent the retransmission limit i.e., a total of $D+1$ transmissions are allowed for each packet before it is dropped. The BSs transmit data packets that are assumed to obey a Poisson arrival process with exogenous arrival rate $\mu$. Since $p$ indicates the network load/ activity factor of the BSs, it is easy to see that $1-p=\exp(-\mu)$. We assume that the net arrival process of the new packets and the retransmitted packets in the steady state is also a Poisson process with arrival rate $\mu_s$ \cite{Dhillon_M2M}. Let $p_s$ denote the activity factor of the BSs at steady state such that $1-p_s=\exp(-\mu_s)$. 

It was shown in \cite{Dhillon_M2M} that  $\mu$ and $\mu_s$ obey the following steady-state relationship
\begin{align}\label{eq:steady_state_arrival_rate}
\mu_s=\mu+\mu_s\epsilon-\mu\delta,
\end{align}
where, $\epsilon$ is the probability that a transmitted packet fails and $\delta=\epsilon^{D+1}$ is the probability that a packet is dropped. Notice that $\epsilon=\mathbb{P}_o$ i.e., the outage probability. Therefore using \eqref{eq:steady_state_arrival_rate}, we have
\begin{align}\label{eq:steady_state_activity}
p_s = 1-(1-p)^{\frac{1-\mathbb{P}_o^{D+1}}{1-\mathbb{P}_o}}.
\end{align}

\begin{figure}
	\centering
	\begin{minipage}{0.48\textwidth}
		\centering
		\includegraphics[width=\linewidth]{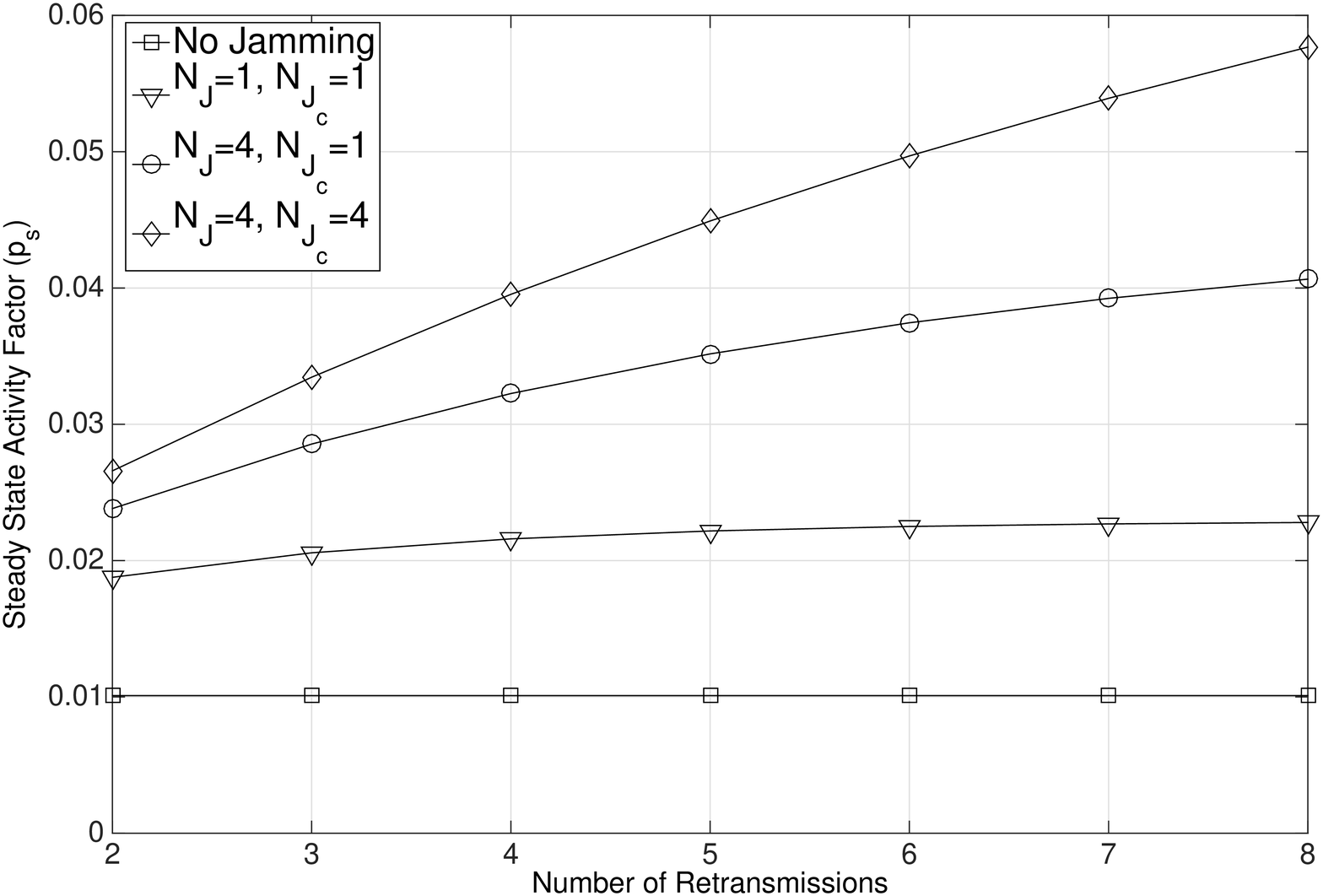}
		\vspace{-20pt}
		\caption{[Effect of Retransmissions]: The steady state activity factor $(p_s)$ as a function of the number of retransmissions $(D)$. The initial activity factor is taken to be $p=0.01$. The $\mathtt{SIR}$ threshold $\theta=0$dB.}
			\label{fig:ReTx1}
	\end{minipage} \hfill
	\begin{minipage}{0.48\textwidth}
		\centering
		\includegraphics[width=\linewidth]{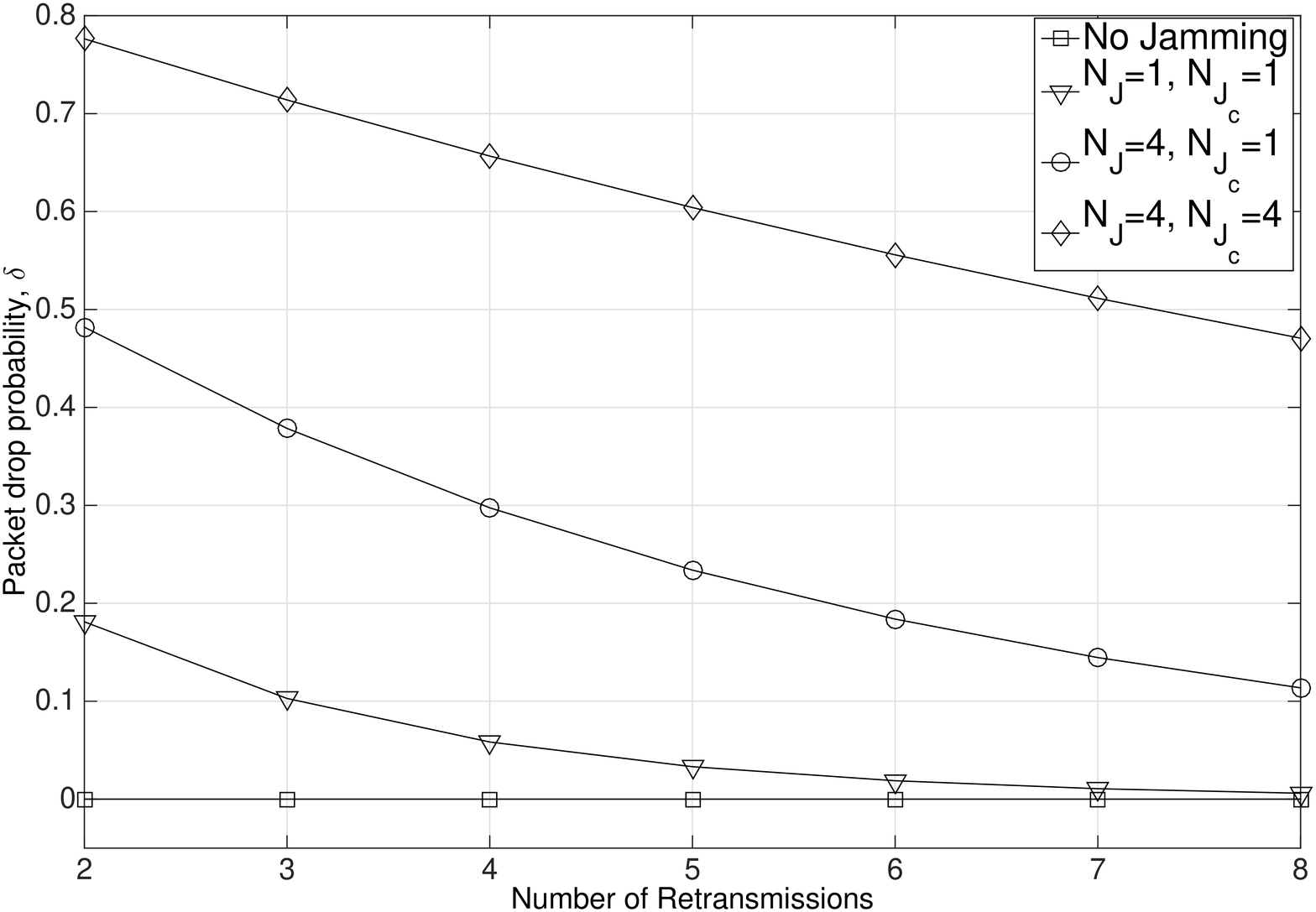}
		\vspace{-20pt}
		\caption{[Effect of Retransmissions]: The steady state packet drop probability $(\delta)$ as a function of the number of retransmissions $(D)$. The initial activity factor is taken to be $p=0.01$. The $\mathtt{SIR}$ threshold $\theta=0$dB.}		
			\label{fig:ReTx2}
	\end{minipage}
\end{figure}

Figs.~\ref{fig:ReTx1}, \ref{fig:ReTx2} show the jamming impact against the wireless network as a function of the number of retransmissions. Specifically, the steady state activity factor $p_s$ is shown as a function of $D$ in Fig.~\ref{fig:ReTx1}. Although $p_s$ increases due to retransmissions, see that this increase is negligible when there is no jamming. However, with only $1$ jammer per cell, the steady state activity factor is doubled when compared to the no-jamming scenario. This indicates that the interference among the BSs increases significantly with only a small number of jammers. The packet dropping probability $\delta$ is shown in Fig.~\ref{fig:ReTx2} as a function of $D$. While $\delta$ decreases when $D$ increases, notice that with $4$ jammers per cell $\delta$ is approximately $50\%$, which suggests that the data cannot be reliably transmitted in the presence of jammers. The results in Figs.~\ref{fig:ReTx1}, \ref{fig:ReTx2} also indicate that due to the increased interference from the BSs (as a result of increasing network load), the number of jammers that would be needed to achieve a required outage probability at the victim receiver can now be decreased. 

\subsection{Error Probability}
As is convention in the wireless communication literature, the power levels considered in the results shown below correspond to transmit $\mathtt{SNR}$ (defined as transmit power/noise power at the victim) and not the received power levels at the victim receiver. Via simulations we observed that for the parameters chosen,  the $\mathtt{SINR}$ at the victim receiver is typically in the range $[-10,30]$dB. 

\begin{figure}
	\centering
	\begin{minipage}{0.48\textwidth}
		\centering
		\includegraphics[width=\linewidth]{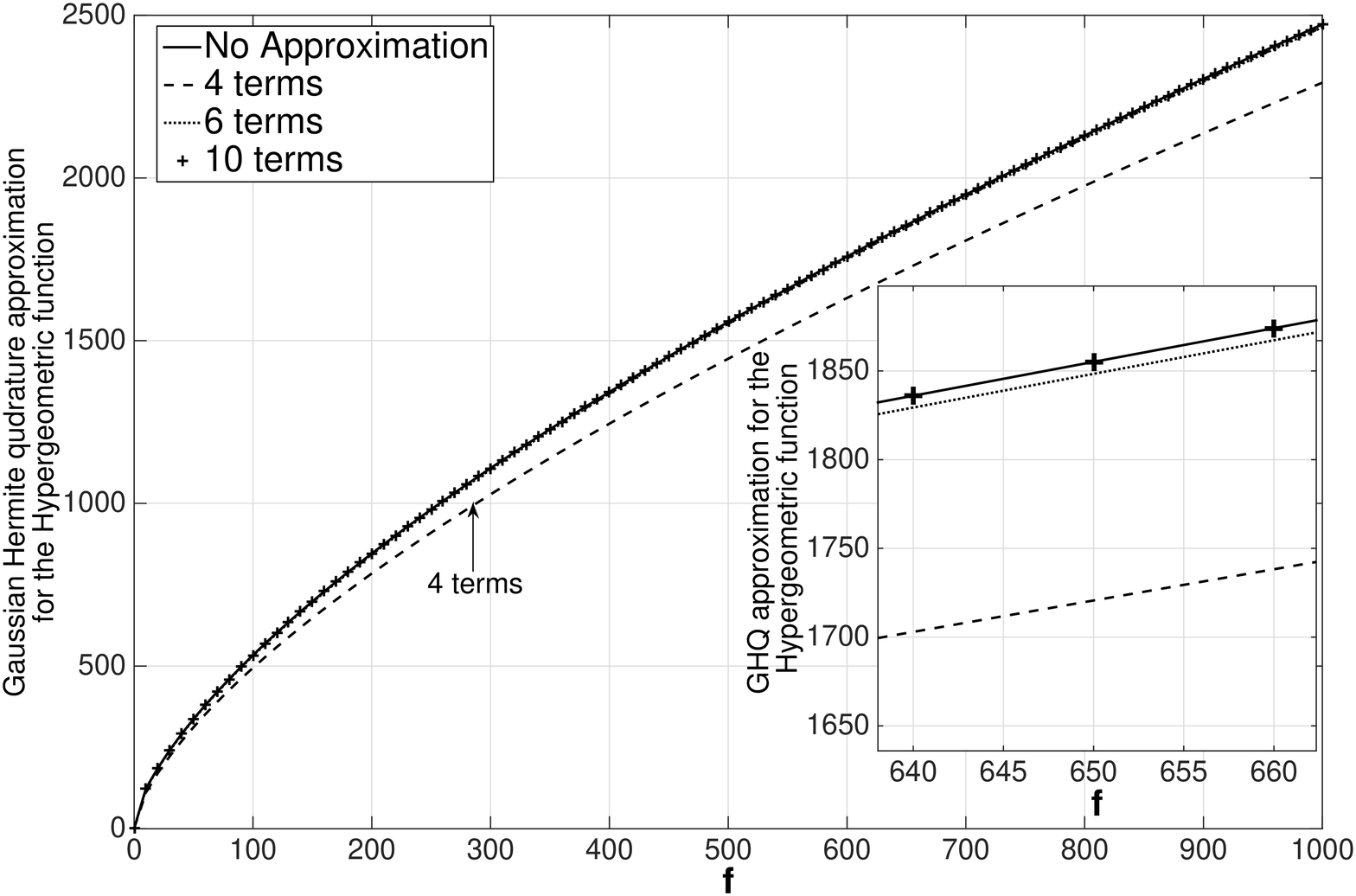}
		\vspace{-20pt}
		\caption{The accuracy of the Gaussian-Hermite quadrature approximation for error probability evaluation as a function of the number of terms $N$ used in the approximation. The zoomed in plot shows a part of the overall figure and indicates that $N=10$ terms very closely matches the true value without any approximation.}
			\label{fig:error_prob_approximation}
	\end{minipage} \hfill
	\begin{minipage}{0.48\textwidth}
		\centering
		\includegraphics[width=\linewidth]{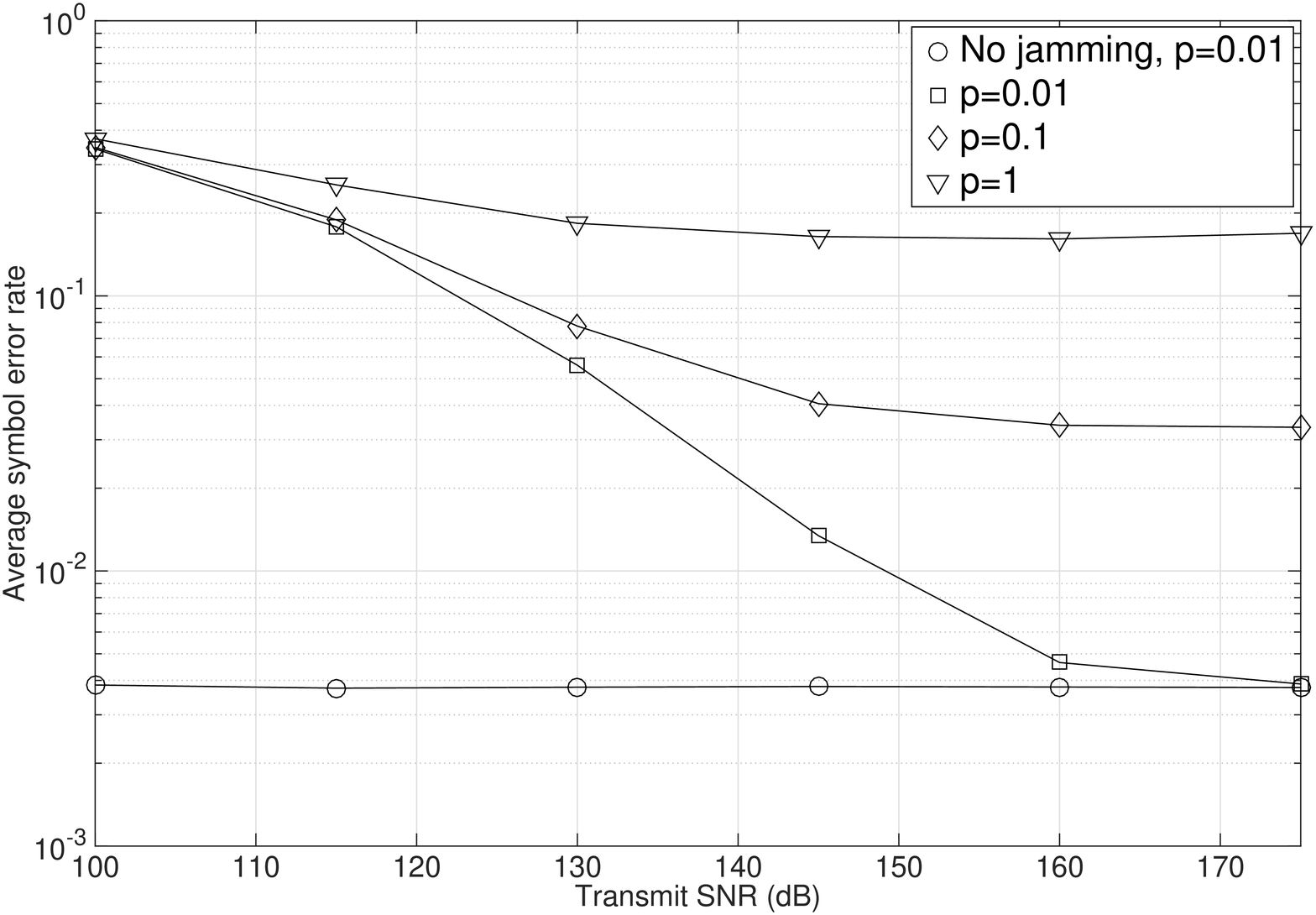}
		\vspace{-20pt}
		\caption{[Effect of Activity Factor]: Average symbol error rate as a function of the activity factor $p$ when the victim receiver uses BPSK modulation and the jammer network uses BPSK modulation. $N_J=4, N_{J_c}=1, \mathtt{JNR}=100$ dB. The solid lines indicate the Monte Carlo simulation results and the markers indicate the theoretical $\mathtt{ASEP}$ evaluated using \eqref{eq:asep1}.}		
			\label{fig_BPSK}
	\end{minipage}
\end{figure}%

\subsubsection{Gaussian-Hermite quadrature approximation to evaluate $\mathbb{E}_{\boldsymbol{\boldsymbol{\Psi}}_J}\left[\Phi_{j_{agg}}(|\omega|)\right]$}
We first discuss the Gaussian-Hermite quadrature approximation used in Lemma~1. We compare \newline $\mathbb{E}_{\chi}\left[{}_1F_1\left[-\frac{1}{\alpha},;1-\frac{1}{\alpha};-f\chi\right]\right]$ and  $\frac{1}{\sqrt{\pi}}\sum_{n=1}^Nw_n{}_1F_1\left[-\frac{1}{\alpha},;1-\frac{1}{\alpha};-f\exp(\sqrt{2}\sigma_{\chi}x_n)\right]$ as a function of $N$ when $\sigma_{\chi}=6$dB. The arguments for the Hypergeometric function are chosen based on the jammer characteristic function shown in Theorem~\ref{theo:jammer_charac_func}. Fig.~\ref{fig:error_prob_approximation} shows the accuracy of the approximation as a function of $N$. It is seen that $N=10$ terms closely matches the true value. Therefore, in what follows we use $N=10$ terms and evaluate the error probability expressions. 


\subsubsection{Effect of number of jammers and activity factor}
\begin{figure}
	\centering
	\begin{minipage}{0.48\textwidth}
		\centering
		\includegraphics[width=\linewidth]{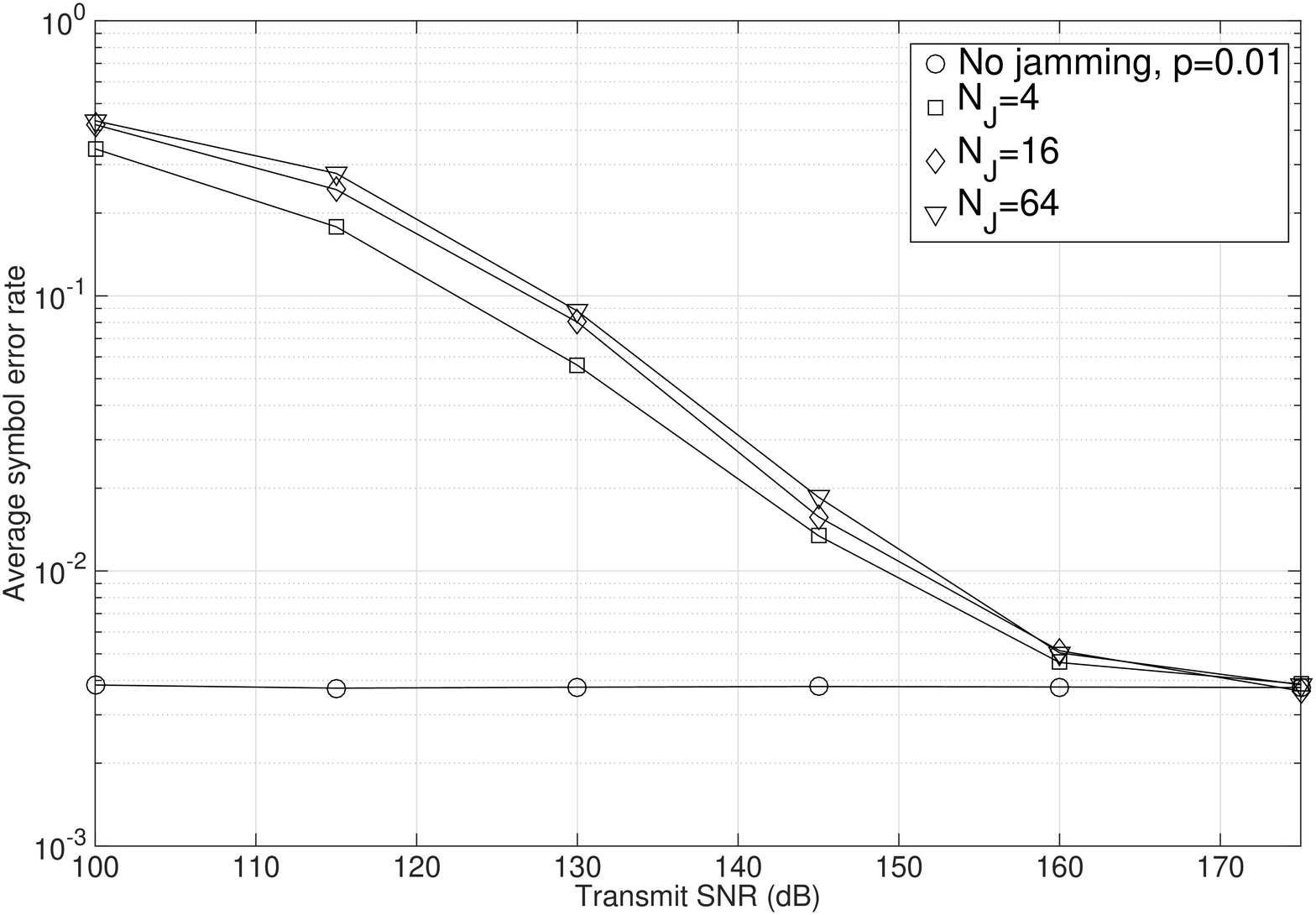}
		\vspace{-20pt}
		\caption{[Effect of Number of Jammers]: Average symbol error rate as a function of the number of jammers when the victim receiver uses BPSK modulation and the jammer network uses BPSK modulation. $N_{J_c}=1$, $p=0.01$. The solid lines indicate the Monte Carlo simulation results and the markers indicate the theoretical $\mathtt{ASEP}$ evaluated using \eqref{eq:asep1}.}
			\label{fig_BPSK_versus_Nj}
	\end{minipage} \hfill
	\begin{minipage}{0.48\textwidth}
		\centering
		\includegraphics[width=\linewidth]{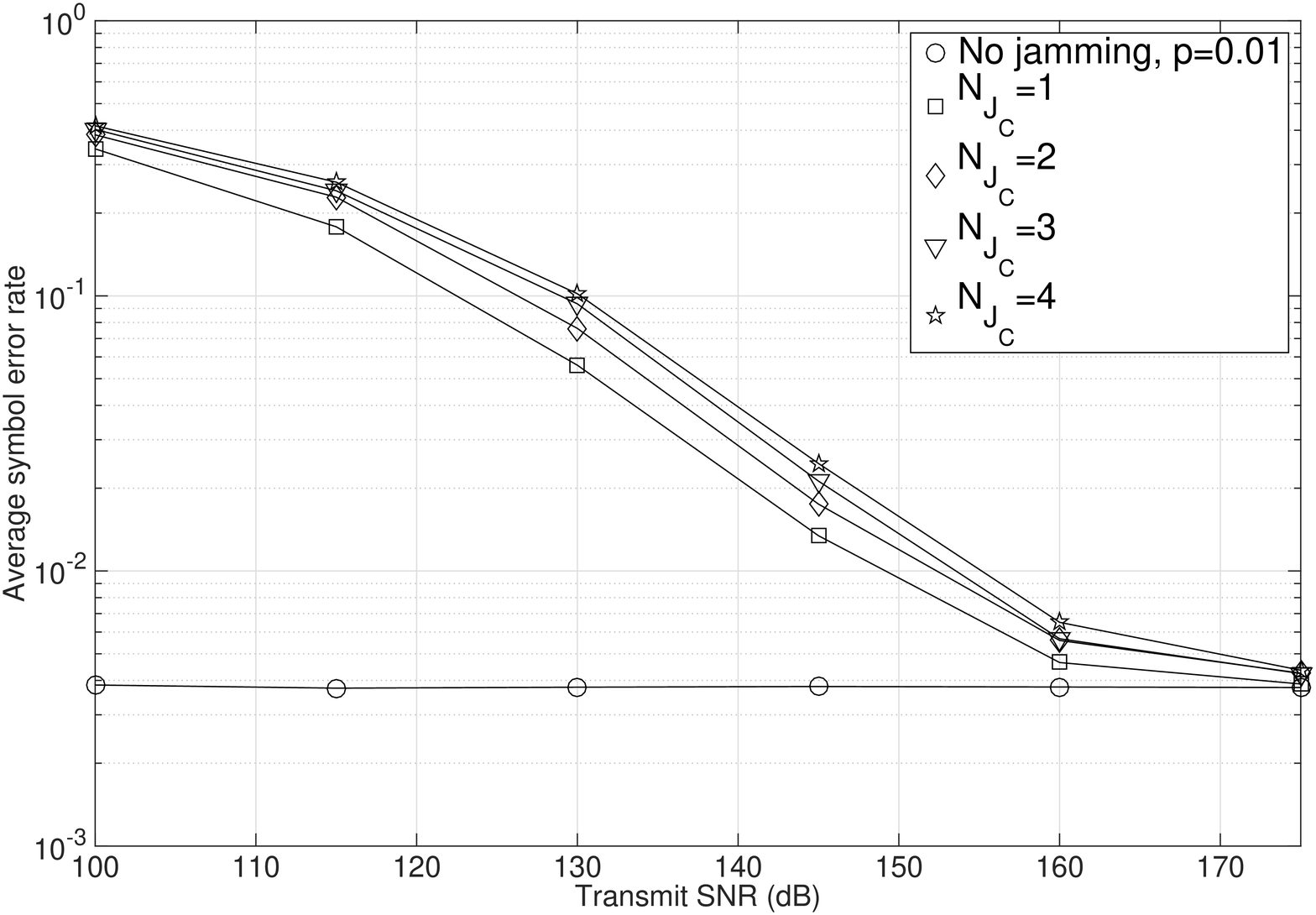}
		\vspace{-20pt}
		\caption{[Effect of $N_{J_c}$]: Average symbol error rate when the victim receiver uses BPSK modulation and the jammer network uses BPSK modulation as a function of the number of jammers per cell (BS). The solid lines indicate the Monte Carlo simulation results and the markers indicate the theoretical $\mathtt{ASEP}$ evaluated using \eqref{eq:asep1}. }		
			\label{fig_BPSK_versus_Njc}
	\end{minipage}
\end{figure}

Fig.~\ref{fig_BPSK} shows the theoretical and the simulation results for the error probability of a victim receiver as a function of the activity factor $p$ when BPSK modulation scheme is used both by the BSs and the jammers. Note that we used BPSK modulation scheme against BPSK victim signal because \cite{ModulationJamming_Journal} indicates that BPSK is the optimal modulation scheme against a BPSK victim signal. However replicating the theoretical analysis in \cite{ModulationJamming_Journal} to the context of networks is beyond the scope of this paper. Instead, we present various simulation results that take into account the various jamming signals that may be used. It is seen in Fig.~\ref{fig_BPSK} that the theoretical $\mathtt{ASEP}$ expressions shown in Section~\ref{sec:ErrorProbability} match perfectly with the Monte Carlo simulation results for various activity factors $p$. Also, $\mathtt{ASEP}$ increases with $p$ due to increased interference from the active interfering BSs. The error probability in a non-jamming scenario is seen to be constant due to the interference-limited scenarios considered in this paper. Fig.~\ref{fig_BPSK_versus_Nj} shows the  theoretical and the simulation results for the victim receiver's error probability as a function of the number of jammers in the network. Notice that the theoretical and the simulation results match perfectly. Also the behavior of $\mathtt{ASEP}$ is as expected-- $\mathtt{ASEP}$ increases with $N_J$ due to increased interference from the jammer network. However, as noted earlier in Fig.~\ref{figPc}, the jamming impact is limited due to path loss as $N_J$ increases because $R_J$ also increases. Fig.~\ref{fig_BPSK_versus_Njc} shows the jamming impact as a function of the number of jammers per cell. It is seen that the error probability of the victim receiver increases with $N_{J_c}$. Note that because we consider interference limited scenarios, the error probability flattens out in all these results. 

\subsubsection{Effect of shadowing}
\begin{figure}
	\centering
	\begin{minipage}{0.48\textwidth}
		\centering
		\includegraphics[width=\linewidth]{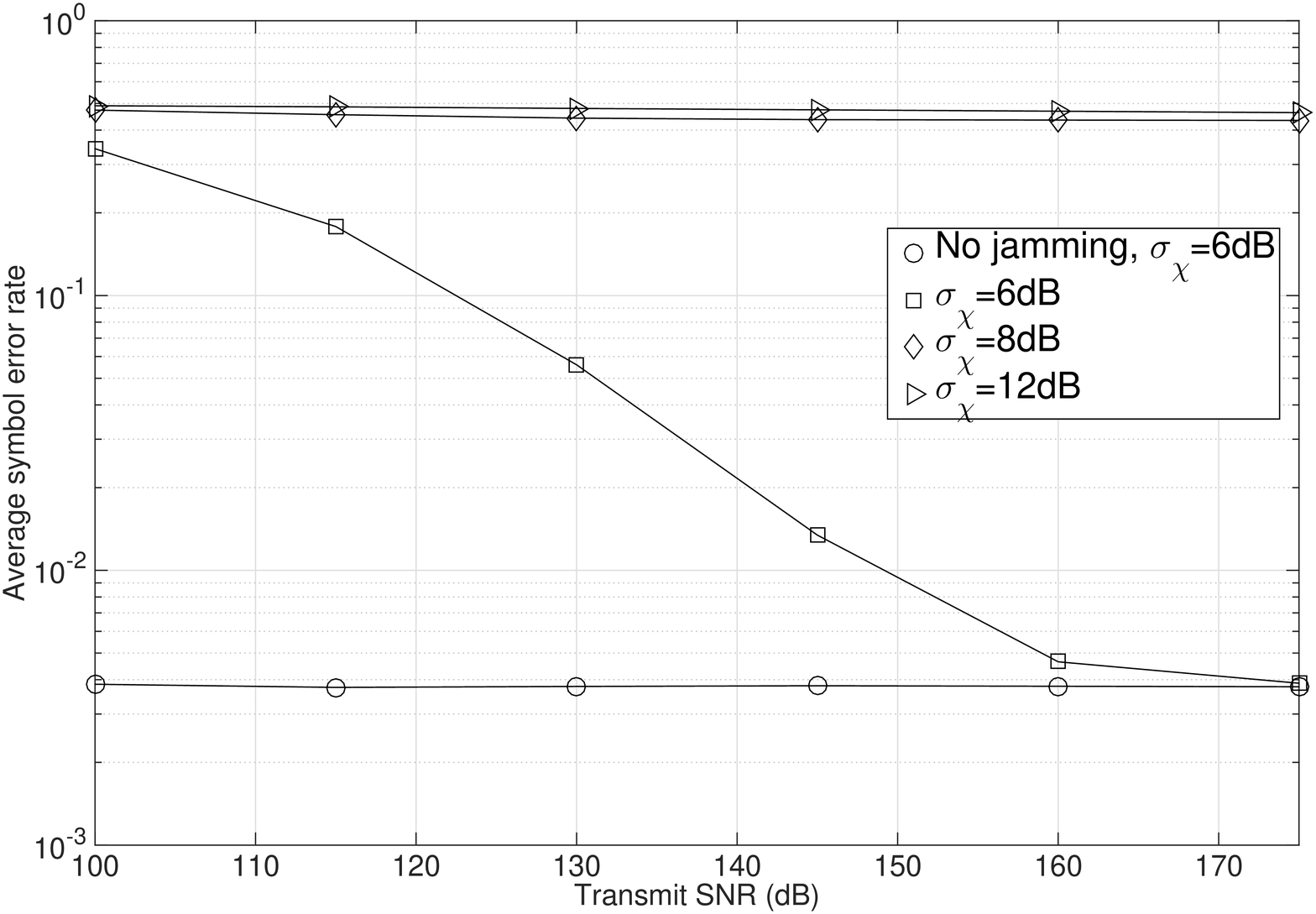}
		\vspace{-20pt}
		\caption{[Effect of shadowing]: Average symbol error rate  as a function of shadowing power level when the victim receiver uses BPSK modulation and the jammer network uses BPSK modulation. The solid lines indicate the Monte Carlo simulation results and the markers indicate the theoretical $\mathtt{ASEP}$ evaluated using \eqref{eq:asep1}.}
			\label{fig_BPSK_versus_shadowingj}
	\end{minipage} \hfill
	\begin{minipage}{0.48\textwidth}
		\centering
		\includegraphics[width=\linewidth]{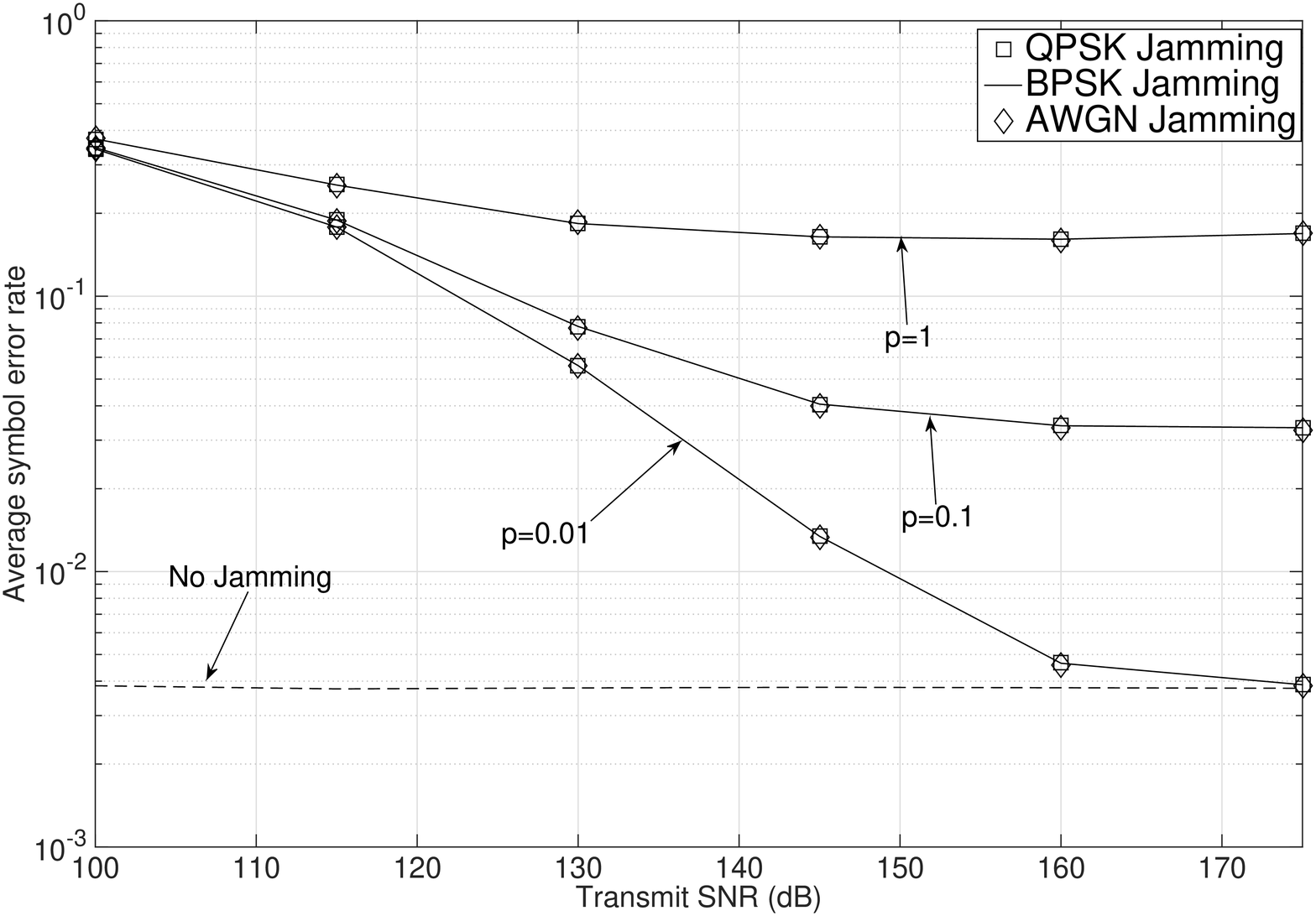}
		\vspace{-20pt}
		\caption{[Effect of the jamming signaling scheme]: Average symbol error rate as a function of $p$ when the victim receiver uses BPSK modulation and different jamming signals are used by the jammer network. $N_J=4$, $N_{J_c}=1$. It is seen that in all cases, the jamming performance of the three jamming signals are the same. }		
			\label{fig_BPSK_Diff_Jamming_Signals}
	\end{minipage}
\end{figure}

Fig.~\ref{fig_BPSK_versus_shadowingj} shows the theoretical and the simulation results for the BPSK modulation scheme used by the BSs and the jammers as a function of $\sigma_{\chi}$. First notice that irrespective of $\sigma_{\chi}$, the theoretical expressions match perfectly with the Monte Carlo simulation results. Further, notice that as seen in the case of the outage probability, the error probability increases with $\sigma_{\chi}$ due to increase in the jammer interference. 

Having established the fact that the theoretical expressions indeed match with the simulation results, we now focus on comparing the performance of the jammer network under a variety of scenarios. 
\subsubsection{Effect of various jamming signals}
Fig.~\ref{fig_BPSK_Diff_Jamming_Signals} shows the jamming behavior of various jamming signals against BPSK modulated victim signals. As was explained earlier, any constant envelope modulation schemes such as BPSK and QPSK will cause the same impact on the victim. Similarly, the AWGN jamming signal will cause the same error rate at the victim as the jammers are not aware of the fading channel between itself and the victim receiver. Therefore, the random channel $g_i$ between the $i$th jammer and the victim receiver randomly rotates the BPSK and QPSK jamming signals which will now appear similar to AWGN signals when they reach the victim receiver. Hence, under such cases the optimal jamming results discussed in \cite{ModulationJamming_Journal} and \cite{FadingJamming_Journal} are not realized. Extending the optimal jamming signaling strategies obtained in \cite{ModulationJamming_Journal} and \cite{FadingJamming_Journal} are beyond the scope of this paper. 

\begin{figure}[ht]
\centering
\vspace{-10pt}
\includegraphics[width=0.55\textwidth]{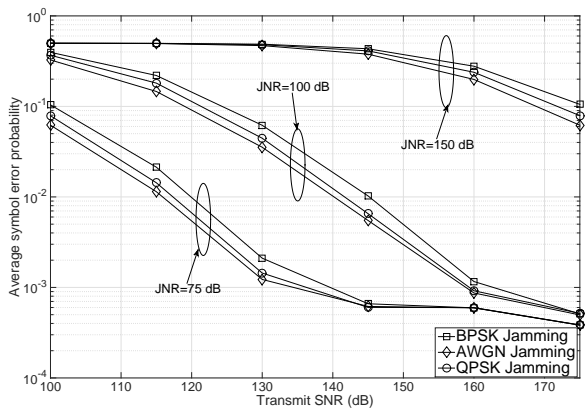}
\vspace{-10pt}
\caption{[No Fading Scenario]: Average symbol error rate when the victim receiver uses BPSK modulation and different jamming signals are used by the jammer network, $N_J=4, N_{J_c}=1$, $p=0.01$. In all cases it is seen that the BPSK jamming outperforms QPSK and AWGN jamming signaling schemes.}
\label{fig_no_fading}
\end{figure}

\subsubsection{No fading channel}
Here, we consider the following system model 
\begin{align}\label{eq:no_fading_system}
y=\sqrt{P_T}(1+r_0)^{-\alpha}s_0+\underbrace{\sum_{i\in\boldsymbol{\boldsymbol{\Psi}}^{(\backslash 0)}}\sqrt{P_T}(1+r_i)^{-\alpha}s_i}_{\boldsymbol{i}_{agg}(r_0)}+\underbrace{\sum_{i\in\boldsymbol{\boldsymbol{\Psi}}_J}\sqrt{P_J\chi^J_i}(1+d_i)^{-\alpha}j_i}_{\boldsymbol{j}_{agg}}+n,
\end{align}
i.e., the small scale fading effects are ignored. Such a model is commonly encountered in line-of-sight-based wireless communication systems. We consider this model to show the best case performance for the jammer network and to show the potential effects that different jamming signaling schemes may have on the victim receiver. Also notice that the analysis presented for both the outage probability and error probability cannot be easily extended to the cases where the fading channel is not considered. 

Fig.~\ref{fig_no_fading} shows the $\mathtt{ASEP}$ for the system model in \eqref{eq:no_fading_system} for various values of $P_J$. First notice that irrespective of the jamming signaling scheme used, the error probability increases with $P_J$. Since fading is not considered in this scenario, notice that the behavior of the various jamming signaling schemes is different which is in contrast to the behavior seen in Fig.~\ref{fig_BPSK_Diff_Jamming_Signals}. Specifically, notice that BPSK jamming outperforms both QPSK and AWGN jamming signaling schemes and that QPSK jamming has a slight advantage over AWGN jamming. BPSK and QPSK jamming outperform AWGN jamming because they are successful in changing the distance between the constellation points of the victim receiver more effectively than the AWGN jamming signal (recall that error probability is typically a function of the minimum distance between the constellation points). This performance of the BPSK jamming is in agreement with the results obtained in \cite{ModulationJamming_Journal} for the point-to-point link scenario. Such a performance analysis can help the BS network to reinforce their signaling strategies to avoid any adversarial attacks. 

\subsection{Limitations and Future Work}
\begin{figure}[ht]
\vspace{-20pt}
\centering
\includegraphics[width=0.55\textwidth]{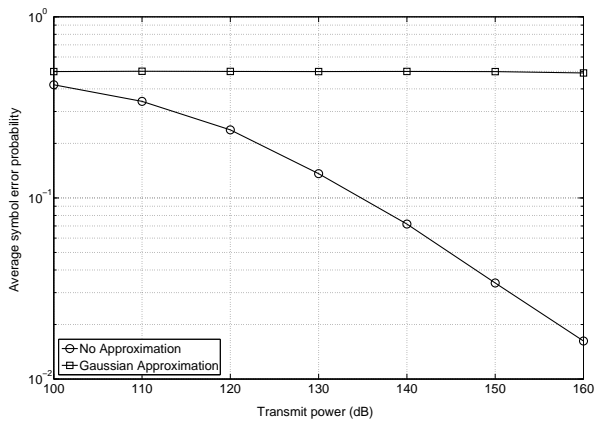}
\vspace{-10pt}
\caption{The symbol error probability of the victim receiver when the jammer interference is approximated as Gaussian with variance denoted by \eqref{eq:Gauss_Approx}.}
\vspace{-10pt}
\label{fig_gauss_Approx}
\end{figure}

\begin{enumerate}
\item As mentioned previously, in interference limited scenarios, the NN approximation (corresponding to the modulation scheme of the victim) gives exact error probability expressions only for the binary modulation schemes. Under the interference limited scenarios, such as the ones studied in this work, this approximation does not accurately evaluate the error probability when higher order modulations are considered. 
\item It is necessary to obtain closed form approximations for the BPP interference so as to employ the EiD technique proposed in \cite{DiRenzo} in order to derive error probability expressions for higher order modulation schemes as well. While the Gaussian approximation for interference modeling has widely been used in the literature, in our analysis we observed that this approximation fails to capture the actual effects of the BPP interference.  Fig.~\ref{fig_gauss_Approx} compares the error probability of the victim receiver when the BPP interference is approximated as Gaussian with mean $0$ and variance given by
\begin{align}\label{eq:Gauss_Approx}
P_JE(\chi^J)\int_{0}^{R_J}(1+r)^{-\alpha}\frac{2r}{R_J^2}\mathrm{d}r=\frac{2P_J\exp\left[\frac{\sigma^2_{\chi}}{2}\right]}{R_J^2(\alpha-2)(\alpha-1)}\left[1-\left[1+R_J\right]^{1-\alpha}\left[1+R_J[\alpha-1]\right]\right]
\end{align}
\item During our study, it was observed that standard computational tools such as Matlab and Mathematica failed in handling singularities in the evaluation of $\mathtt{ASEP}$ shown in \eqref{eq:asep1}. Despite using the approximations for the hypergeometric functions suggested in \cite{DiRenzo_TCOM}, we observed that these tools provided significantly different results especially when handling higher order modulations such as QPSK and $16$-QAM. It is therefore necessary to find alternative techniques that enable us to analyze higher order modulation schemes as well. 
\end{enumerate}
\vspace{-10pt}
\section{Conclusion}\label{sec:Conclusion}
In this paper, we studied jamming against wireless networks from a physical layer perspective by employing tools from stochastic geometry. Specifically, we studied jamming against a network of BS/AP that are increasingly being modeled according to a PPP. Since the victim locations are typically not known \emph{a priori} we modeled the jammer network according to a BPP and studied the wireless network performance using the outage and error probability metrics of a victim receiver. Several interesting insights about the jammer network behavior were discussed in terms of these two metrics. We showed that the number of jammers required to attain a specific outage probability increases (decreases) with victim signal power (jammer power and network loading). Specifically, it was seen that with only $1$ jammer per BS/AP, the outage probability of the wireless network can be increased from $1\%$ (in the non-jamming case) to $80\%$ and when retransmissions are used, the effective network activity factor (interference among the BSs) can be doubled. It was observed that the behavior with respect to the BS density is not obvious. Specifically, we showed that the required number of jammers increases with the density of the BS nodes $\lambda_T$ only until a certain point beyond which it decreases. Furthermore, we also analyzed the error probability of the victim receiver, both from a simulation and a theoretical perspective and showed that the exact error probability expressions can be evaluated in the case of binary modulations. We showed that some recent results related to modulation-based jamming in a point-to-point link setting cannot be directly extended to the case of jamming against wireless networks. 

\section*{Appendix A - Proof of Theorem 1}
The outage probability expression of the victim receiver under attack by the BPP distributed jammer network is given by
\begin{align}\label{BPP_Pc_Appendix}
P_o(\mathtt{SIR}<\theta)&=1-P\left[\frac{P_T |h_0|^2 (1+r_0)^{-\alpha}}{\sum_{i \in \boldsymbol{\boldsymbol{\Psi}}\backslash{0}}a_iP_T|h_i|^2(1+r_i)^{-\alpha}+\sum_{i\in \boldsymbol{\boldsymbol{\Psi}}_J}P_J\chi^J_i|g_i|^2(1+d_i)^{-\alpha}}>\theta\right]\nonumber \\
&=1-P\left[|h_0|^2>\theta\frac{\sum_{i \in \boldsymbol{\boldsymbol{\Psi}}\backslash{0}}a_i|h_i|^2(1+r_i)^{-\alpha}}{(1+r_0)^{-\alpha}}+\theta\frac{\sum_{i\in \boldsymbol{\boldsymbol{\Psi}}_J}P_J\chi^J_i|g_i|^2(1+d_i)^{-\alpha}}{P_T(1+r_0)^{-\alpha}}\right]\nonumber
\end{align}
\begin{align}
&\stackrel{(i)}{=}1-\mathbb{E}\left[\exp\left[-\left[\theta\frac{\sum_{i \in \boldsymbol{\boldsymbol{\Psi}}\backslash{0}}a_i|h_i|^2(1+r_i)^{-\alpha}}{(1+r_0)^{-\alpha}}+\theta\frac{\sum_{i\in \boldsymbol{\boldsymbol{\Psi}}_J}P_J\chi^J_i|g_i|^2(1+d_i)^{-\alpha}}{P_T(1+r_0)^{-\alpha}}\right]\right]\right] \nonumber \\
&\stackrel{(ii)}{=}1-\int_{r_0=0}^{\infty}\mathbb{E}\left[\exp\left[-\left[\theta\frac{\sum_{i \in \boldsymbol{\boldsymbol{\Psi}}\backslash{0}}a_i|h_i|^2(1+r_i)^{-\alpha}}{(1+r_0)^{-\alpha}}\right]\right]\right]\times \nonumber \\
&\hspace{50pt}\mathbb{E}\left[\exp\left[-\left[\theta\frac{\sum_{i\in \boldsymbol{\boldsymbol{\Psi}}_J}P_J\chi^J_i|g_i|^2(1+d_i)^{-\alpha}}{P_T(1+r_0)^{-\alpha}}\right]\right]\right]2\pi\lambda_Tr_0\exp(-\lambda_T \pi r_0^2)\mathrm{d} r_0 \nonumber \\
&\stackrel{(iii)}{=}1-\int_{r_0=0}^{\infty}\exp\left[-2\pi p\lambda_T\int_{r=r_0}^{\infty}\left[1-\frac{1}{1+\theta (1+r_0)^{\alpha}(1+r)^{-\alpha}}\right]r\mathrm{d}r\right]\times\nonumber
\end{align}
\begin{align}
&\hspace{35pt}\left[\frac{2}{R_J^2}\int_{r=0}^{R_J}\int_{\chi=0}^{\infty}\frac{1}{1+\frac{\theta (1+r_0)^{\alpha}P_J\chi}{P_T}(1+r)^{-\alpha}}\frac{r\exp\left[-\left[\frac{\log(\chi)}{\sqrt{2\sigma_{\chi}}}\right]^2\right]}{\chi\sqrt{2\sigma_{\chi}}}\mathrm{d}\chi\mathrm{d}r\right]^{N_J} \times \nonumber \\
&\hspace{35pt}2\pi\lambda_Tr_0\exp(-\lambda_T \pi r_0^2)\mathrm{d} r_0
\end{align}
where $(i)$ follows from the fact that $|h_{0}|^2$ is an exponential random variable with mean $1$, $(ii)$ follows from the fact that the pdf of $r_0$ is $2\pi\lambda_Tr_0\exp(-\lambda_T\pi r_0^2)$, and $(iii)$ follows by using the probability generating functional for the PPP \cite{PPP_Dhillon}, the moment generating function for the BPP \cite{Haenggi_BPP}, the fact that $\mathbb{E}(a_i)=p$, $|h_i|^2, |g_i|^2$ are independent unit mean exponential random variables, and $\chi^J_i$ are independent log-normal random variables with mean $0$ and variance $\sigma_{\chi}$.

\section*{Appendix B - Proof of Theorem~2}
The jammer's characteristic function is evaluated as
\begin{align}\label{phi_J_inter}
&\mathbb{E}_{\boldsymbol{\boldsymbol{\Psi}}_J}\left[\Phi_{j_{agg}}(|\omega|)\right]=E_{\boldsymbol{\boldsymbol{\Psi}}_{j}}\left[\prod_{i=1}^{N_J}\mathbb{E}_{\hat{z}}\left[\cos\left[|\omega|\sqrt{P_J}(1+d_i)^{-\frac{\alpha}{2}}\hat{z}\right]\right]\right] \nonumber \\
&\stackrel{(i)}{=}\left[E_{\boldsymbol{\boldsymbol{\Psi}}_{j}}\mathbb{E}_{\hat{z}}\left[\cos(|\omega|\sqrt{P_J}(1+d_i)^{-\frac{\alpha}{2}}\hat{z})\right]\right]^{N_J} \stackrel{(ii)}{=}\left[\mathbb{E}_{\hat{z}}\left[\int_{0}^{R_J}\cos(|\omega|\sqrt{P_J}(1+r)^{-\frac{\alpha}{2}}\hat{z})\frac{2r}{R_J^2}\mathrm{d}r\right]\right]^{N_J} \nonumber \\
&=\left[\mathbb{E}_{\hat{z}}\left[\int_{|\omega|\sqrt{P_J}(1+R_J)^{-\frac{\alpha}{2}}}^{|\omega|\sqrt{P_J}}\cos(t\hat{z})\frac{4}{R_J^2t\alpha}\left[\frac{|\omega|\sqrt{P_J}}{t}\right]^{\frac{2}{\alpha}}\left[\left[\frac{|\omega|\sqrt{P_J}}{t}\right]^{\frac{2}{\alpha}}-1\right]\mathrm{d}t\right]\right]^{N_J},
\end{align}
where $(i)$ and $(ii)$ follow because the jammers and their locations are independent and identically distributed. We now state two results that are used in simplifying the above expression. 
 
The first result is \cite[Eq. 3.771.4]{Integrals_Book}
\begin{align}\label{identity_1}
\int_{0}^{u}\left[1-\cos(tz)\right]t^{-(1+2/c)}dt=\frac{cu^{-2/c}}{2}\left[-1+{}_1F_2\left[-\frac{1}{c},\left[\frac{1}{2},1-\frac{1}{c}\right],-\frac{u^2z^2}{4}\right]\right], \mathfrak{R}\left(\frac{1}{c}\right)<1.
\end{align}
Next, we evaluate $\mathbb{E}_{\hat{z}}(\hat{z}^{2q})$. Recall $\hat{z}=\mathfrak{R}(z)=\sqrt{\chi}|g_i||j_i|\cos(\angle{g_i}+\angle{j_i})$. First notice that $|g_i|\cos(\angle{g_i}+\angle{j_i})$ is a Gaussian random variable with mean $0$ and variance $\frac{1}{2}$ (because $g_i$ is a complex Gaussian random variable). 
Then by using \eqref{identity_2_pre}, we have
\begin{align}\label{identity_2}
\mathbb{E}_{\hat{z}}(\hat{z}^{2q})
&=\frac{\Gamma(q+\frac{1}{2})\mathbb{E}(\chi^q)}{|\mathcal{M}_J|\sqrt{\pi}}\sum_{j_i\in \mathcal{M}_J}|j_i|^{2q},
\end{align}
Substituting \eqref{identity_1} and \eqref{identity_2} in \eqref{phi_J_inter}, using the series expansion of hypergeometric function in \eqref{eq:hyp_geom_series} and noting that $(\frac{1}{2})_q=\frac{\Gamma(q+\frac{1}{2})}{\sqrt{\Pi}}$ and ${}_1F_2[a;b,1;x]={}_1F_1[a;b;x]$, we have \eqref{eq:jammer_charac_func}. 
Note that by using Lemma~1, the expectation of the Hypergeometric functions with respect to $\chi$ can be approximated by a summation of $N$ terms. 
%
\section*{Appendix C - Proof of Corollary~\ref{corr:poiss_char_func}} 
Here, we evaluate the characteristic function of the interference from the BSs other than the serving BS, $\Phi_{i_{agg}}(\omega;r_0)$ where $i_{agg}(r_0)=\sum_{i\in\boldsymbol{\boldsymbol{\Psi}}\backslash \{0\}}\sqrt{P_T\chi_i}(1+r_i)^{-\frac{\alpha}{2}}h_is_i$. Since the effects of shadowing have been considered by transforming the original PPP, we only need to consider $i_{agg}(r_0)=\sum_{i\in\boldsymbol{\boldsymbol{\Psi}}\backslash \{0\}}\sqrt{P_T}(1+r_i)^{-\frac{\alpha}{2}}h_is_i$ to evaluate the BS interference. We have
\begin{align}
\Phi_{i_{agg}}(\omega;r_0) &= \mathbb{E}_{\boldsymbol{\boldsymbol{\Psi}}\backslash \{0\}}\left[\exp\left[j\omega\sum_{i\in\boldsymbol{\boldsymbol{\Psi}}\backslash \{0\}}\sqrt{P_T}(1+r_i)^{-\frac{\alpha}{2}}h_is_i\right]\right] \nonumber \\
&=\mathbb{E}_{\boldsymbol{\boldsymbol{\Psi}}\backslash \{0\}}\left[\prod_{i\in\boldsymbol{\boldsymbol{\Psi}}\backslash \{0\}}\exp\left[j\omega\sqrt{P_T}(1+r_i)^{-\frac{\alpha}{2}}h_is_i\right]\right]\nonumber \\
&\stackrel{(i)}{=}\exp\left[2\pi p\lambda_T\int_{r_0}^{\infty}\left[\mathbb{E}\left[\exp\left[j\omega\sqrt{P_T}(1+r_i)^{-\frac{\alpha}{2}}h_is_i\right]\right]-1\right]r\mathrm{d}r\right] \nonumber \\
&\stackrel{(ii)}{=}\exp\left[2\pi p\lambda_T\int_{r_0}^{\infty}\left[\Phi_z\left[\omega\sqrt{P_T}(1+r_i)^{-\frac{\alpha}{2}}\right]-1\right]r\mathrm{d}r\right],
\end{align}
where $(i)$ follows by using the probability generating functional of the PPP \cite{PPP_Dhillon} and $(ii)$ follows by defining a new variable $z=h_is_i$ which is a circularly symmetric random variable. Since the interfering BSs are active with a probability $p$, they form a thinned PPP with density $p\lambda_T$ \cite{DhillonActivityFactor} and therefore the effect of $p$ is seen in the characteristic function of $i_{agg}(r_0)$. In other words, we observe a thinning of the interference field \cite{DhillonActivityFactor}. Now notice that the expression inside the integral above is similar to what was seen in the derivation of $\mathbb{E}_{\Phi_{j_{agg}}}(\Phi_{j_{agg}}(\omega))$ in \eqref{eq:jam_intf_charac_initial}. Therefore, by following similar steps as in the case of jammer interference in \eqref{eq:jam_intf_charac_initial} and the analysis shown in the evaluation of \eqref{phi_J_inter}, the overall BS interference is given by \eqref{eq:poiss_char_func}.
\end{document}